\documentclass[lettersize,journal]{IEEEtran}
\usepackage{amsmath,amsfonts}
\usepackage{amssymb}
\usepackage{algorithmic}
\usepackage{algorithm}
\usepackage{array}
\usepackage[caption=false,font=normalsize,labelfont=sf,textfont=sf]{subfig}
\usepackage{textcomp}
\usepackage{stfloats}
\usepackage{url}
\usepackage{verbatim}
\usepackage{graphicx}
\usepackage{cite}
\usepackage{booktabs}
\usepackage{xcolor}
\usepackage{bm}

\begin{document}

\title{Synergistic Localization and Sensing in MIMO-OFDM Systems via Mixed-Integer Bilevel Learning}

\author{Zelin Zhu,~\IEEEmembership{}
       Kai Yang,~\IEEEmembership{Senior Member,~IEEE,}
       Rui Zhang,~\IEEEmembership{Fellow,~IEEE}

\thanks{This paper was produced by the IEEE Publication Technology Group. They are in Piscataway, NJ.}
\thanks{Zelin Zhu and Kai Yang (Corresponding Author) are with the Department of Computer Science and Technology, Tongji University, ShangHai 200092, China (e-mail: kaiyang@tongji.edu.cn). R. Zhang is with the Chinese University of Hong Kong (Shenzhen), Guangdong 518172, China. He is also with the Department of Electrical and Computer Engineering, National University of Singapore, Singapore 117583 (e-mail: elezhang@nus.edu.sg).}}


\maketitle

\begin{abstract}
Wireless localization and sensing technologies are essential in modern wireless networks, supporting applications in smart cities, the Internet of Things (IoT), and autonomous systems. High-performance localization and sensing systems are critical for both network efficiency and emerging intelligent applications. Integrating channel state information (CSI) with deep learning has recently emerged as a promising solution. Recent works have leveraged the spatial diversity of multiple input multiple output (MIMO) systems and the frequency granularity of orthogonal frequency division multiplexing (OFDM) waveforms to improve spatial resolution. Nevertheless, the joint modeling of localization and sensing under the high-dimensional CSI characteristics of MIMO-OFDM systems remains insufficiently investigated. This work aims to jointly model and optimize localization and sensing tasks to harness their potential synergy. We first formulate localization and sensing as a mixed-integer bilevel deep learning problem and then propose a novel stochastic proximal gradient-based mixed-integer bilevel optimization (SPG-MIBO) algorithm. SPG-MIBO is well-suited for high-dimensional and large-scale datasets, leveraging mini-batch training at each step for computational and memory efficiency. The algorithm is also supported by theoretical convergence guarantees. Extensive experiments on multiple datasets validate its effectiveness and highlight the performance gains from joint localization and sensing optimization.
\end{abstract}

\begin{IEEEkeywords}
Wirelesss localization and sensing, Bilevel machine learning, Mixed-integer bilevel optimization.
\end{IEEEkeywords}

\section{Introduction}
\begin{figure}[htbp]
\centering
\includegraphics[width=\columnwidth,trim=90 0 50 0, clip]{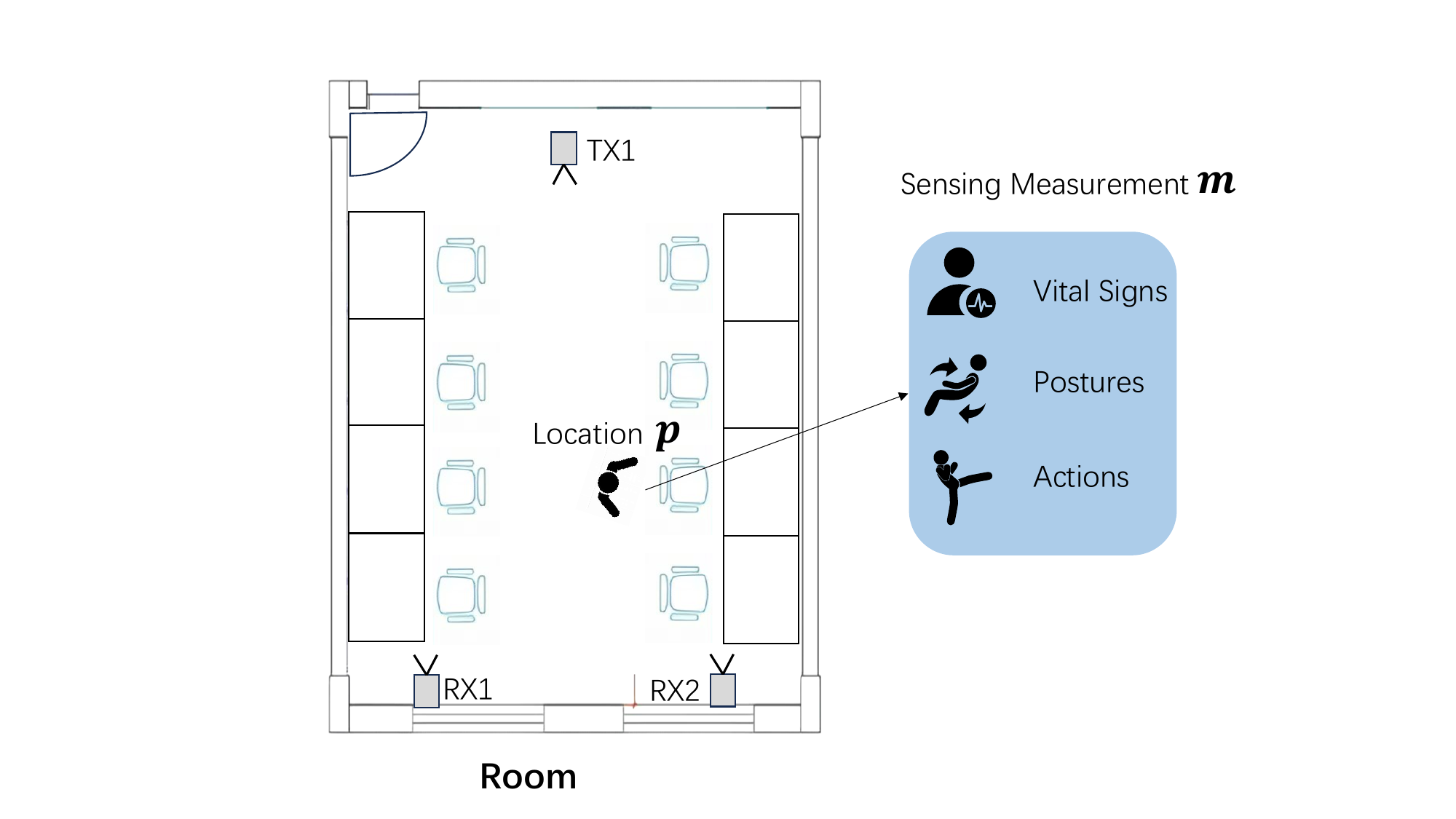} 
\caption{Joint Wireless Localization and Sensing Scenario}
\label{Joint scenario}
\end{figure}

\IEEEPARstart{W}{ireless} localization and sensing are fundamental to modern wireless systems, especially in Integrated Sensing and Communication (ISAC) networks \cite{li2023joint} and smart home technologies \cite{montoliu2020indoor}. Applications such as remote healthcare \cite{bao2022wi}, intrusion detection \cite{jin2018whole}, and virtual reality \cite{fu2018writing} rely heavily on accurate indoor localization and sensing. Among various technologies, Wi-Fi-based passive localization and sensing have attracted extensive research attention due to Wi-Fi’s ubiquity. These methods typically exploit wireless channel information, such as Received Signal Strength Indicator (RSSI) or Channel State Information (CSI), to infer target position or state (see Fig.\ref{Joint scenario}).

Traditional model-based approaches \cite{liu2015rss, gu2019wifi, wang2020csi} build physical models to relate target states to wireless signals, but they often rely on ideal assumptions that are rarely met in real scenarios. Recently, deep learning methods \cite{xie2024robust, kontou2023contactless, ngamakeur2023passive} have gained prominence by learning complex features from large datasets, significantly improving localization and sensing accuracy and robustness.

Practically, in Orthogonal Frequency Division Multiplexing (OFDM)-based wireless systems, there exists a subset of frequency subcarriers that are highly responsive to target-induced signal variations and thus beneficial for both localization and sensing tasks. Conversely, some subcarriers exhibit low sensitivity and may contribute little to either task, potentially degrading their overall model performance. This phenomenon becomes particularly pronounced in distributed Multiple-Input Multiple-Output (MIMO) wireless systems, where certain transmitter/receiver (TX/RX) pairs may have limited coverage or weak signal interaction with the target area, causing their corresponding subcarriers to show minimal response to target activity. However, most existing deep leaning- based works treat localization and sensing as separate tasks, ignoring their shared wireless signal source and inherent correlation \cite{yan2021device}. Inspired by work on wireless localization \cite{wang2018low, wang2018quasi} and sensing \cite{zhang2022hybrid, li2023joint} where the performance of enhancing localization and sensing models is achieved by extracting discriminative information from high-dimensional subcarriers. These prior works highlights the importance of identifying and leveraging the most informative subcarriers for joint localization and sensing. Thus, designing a shared subcarrier selection module that filters out subcarriers insensitive to target-induced variations can enhance feature extraction for both tasks.

In addition, localization can theoretically be regarded as a coarse-grained form of sensing. While sensing aims to capture fine-grained target states such as gestures, postures, or micro-movements, localization focuses on estimating the spatial position of the target, which provides a more abstract and general description of its state. In joint modeling, this hierarchical relationship implies that accurate localization can serve as a prior or contextual constraint to support more detailed sensing tasks. Therefore, treating localization as a coarse-grained sensing process not only aligns with the shared signal basis but also facilitates the construction of integrated localization and sensing with a hierarchical structure. This motivates modeling the joint localization and sensing task as a bilevel optimization problem. 

Bilevel optimization has been widely studied in machine learning \cite{liu2021investigating, zhang2022revisiting, chen2022gradient}, particularly for problems with hierarchical structures, such as, meta-learning, where meta-parameters and model parameters form two-level optimization tasks. Unlike most existing bilevel machine learning studies that focus mainly on continuous variables, the synergistic fusion problem of localization and sensing involves mixed-integer variables to represent discrete subcarrier weights in Multiple-Input Multiple-Output and Orthogonal Frequency Division Multiplexing (MIMO-OFDM) wireless signals, reflecting the inherent discrete decisions in wireless systems. Such mixed-integer bilevel problems are highly challenging to solve because of their non-convexity, scale, and combinatorial nature. Although there exist some bilevel optimization works\cite{miao2016genetic,tahernejad2020branch,tang2016class,avraamidou2017mixed} that consider the mixed-integer setting, their applicability to large-scale parameter deep learning is limited due to the constraints on problem types and the reliance on methods with sensitivity of computational complexity to parameter dimensionality, such as heuristic methods and genetic methods.

Gradient-based methods\cite{dumouchelle2024neur2bilo,jiao2022asynchronous,jiao2022distributed,huang2024triadic,chen2024robust,jian2024tri} have attracted significant attention in bilevel machine learning due to their computational efficiency, making them well-suited for large-scale optimization. In particular, stochastic gradient methods enable effective optimization on large-sample datasets through mini-batch training, which is crucial for scalable deep model learning. Therefore, developing a Mixed-Integer Bilevel Optimization (MIBO) algorithm based on stochastic gradients is both necessary and practical for handling high-dimensional problems in real-world wireless sensing and localization scenarios.

Motivated by the above discussions and the identified gaps in prior work, we summarize the main contributions of our current work as follows:
\begin{itemize}
    \item To the best of our knowledge, this work is the first to formulate the joint localization and sensing problem within the framework of mixed-integer bilevel learning, thereby introducing a novel perspective that captures both the discrete and continuous aspects of wireless system modeling. Our approach explicitly exploits the hierarchical structure inherent in localization and sensing, where task interdependencies can be modeled at different levels of abstraction. This unified formulation not only bridges the gap between these two closely related tasks but also provides a principled way to harness their mutual benefits through joint optimization.
    \item We design a Stochastic Proximal Gradient-based Mixed-Interger Bilevel Optimization (SPG-MIBO) algorithm to effectively address the challenges of mixed-integer bilevel optimization problems with binary constraints. The algorithm is carefully constructed to ensure computational efficiency, while its stochastic proximal gradient-based structure allows it to scale seamlessly to high-dimensional data scenarios induced by MIMO-OFDM systems. Furthermore, the algorithm maintains memory efficiency, making it well-suited for deployment in real-world systems with limited computational resources. 
    \item The convergence rate of the proposed SPG-MIBO algorithm is rigorously derived, providing strong theoretical support for its practical deployment. To further demonstrate its effectiveness and generalizability, we conduct extensive experiments on multiple real-world and synthetic datasets. The experimental results consistently validate the advantages of the proposed method, highlighting its performance benefits in jointly optimizing localization and sensing tasks.
\end{itemize}

\section{Related Works}
\subsection{Wireless Localization and Sensing}
Recent advances in wireless sensing and localization have moved from using low-dimensional features such as RSSI \cite{yang2013rssi} to exploiting fine-grained CSI \cite{wu2022wifi}. Multi-subcarrier CSI provides high spatial and frequency resolution, enabling fine-grained applications ranging from localization to human activity recognition \cite{jin2018whole, yan2019wiact}, and even vital sign monitoring \cite{sharma2022attention, fan2024contactless}. System performance can be further enhanced by using rotating antennas \cite{graefenstein2009wireless} or distributed multi-TX/RX configurations \cite{matricardi2023performance}.

Although both sensing and localization rely on the same wireless channel observations, most existing works treat them as isolated tasks. Some efforts have attempted sequential or loosely coupled frameworks \cite{yan2021device}, while recent deep learning approaches \cite{zhang2022wi} have proposed early fusion using shared feature representations. However, these approaches still face challenges due to the high dimensionality of CSI, which complicates feature extraction and model training.

To address the challenges posed by high-dimensional CSI data in MIMO-OFDM systems, subcarrier selection has been proposed to reduce redundant or uninformative signals, thereby improving the efficiency and robustness of either localization \cite{wang2018low} or sensing \cite{zhang2022hybrid, li2023joint}. Works related to subcarrier selection \cite{patwari2013breathfinding,hillyard2018experience,gu2018your ,wang2020resilient} find that getting the subcarriers that are more relevant to the target reduces multi-path interference from the physical perspective. However, existing studies typically apply subcarrier selection to a single task and do not consider it in a unified framework. 

In addition, despite the potential of OFDM-based and distributed multi-TX/RX systems to provide diverse spatial and frequency observations\cite{fahad2025ensemble,chu2021wifi}, there has been little effort to systematically exploit this structure via hierarchical modeling. Specifically, no existing work has investigated a bilevel optimization approach that simultaneously achieves dimensionality reduction through subcarrier selection and facilitates multitask learning in such distributed wireless settings. 

Furthermore, the use of integer variables to explicitly enforce subcarrier selection constraints, thereby enhancing interpretability and sparsity has been largely overlooked in learning-based frameworks. These gaps motivate our formulation of the joint localization and sensing problem as a mixed-integer bilevel optimization task that can naturally accommodate discrete structure, shared representation, and scalable training via stochastic gradients.

\subsection{Bilevel Machine Learning and Mixed-Integer Bilevel Optimization}
Bilevel optimization has been widely studied in machine learning \cite{liu2021investigating, biswas2019literature, chen2022gradient}, especially for problems with hierarchical structure, such as meta-learning. Gradient-based bilevel solvers have shown strong scalability for large models with continuous variables\cite{chen2022gradient}.

However, handling discrete variables, such as  \{0,1\} integer parameters in subcarrier selection remains a major challenge. Classical mixed-integer bilevel optimization (MIBO) methods, such as branch-and-bound \cite{tahernejad2020branch} or heuristic-based approaches \cite{miao2016genetic, denegre2011interdiction, soares2023deterministic}, often lack scalability or convergence guarantees. While \cite{dumouchelle2024neur2bilo} proposes a neural solver for MIBO with large-scale parameters, it lacks support for stochastic gradients, limiting its application to deep models trained on large datasets (see in TABLE \ref{tab:MIBO methods}).

Recent works \cite{beykal2020domino} introduce gray-box or surrogate modeling for MIBO in learning, but it trades off interpretability or theoretical guarantees. Thus, there remains a critical need for scalable, gradient-based MIBO algorithms that can jointly optimize discrete structure and continuous parameters under large-scale data settings.

\begin{table*}[htb]
\caption{Mixed-Integer Bilevel Optimization Methods Overview}
\centering
\resizebox{0.7\textwidth}{!}{
\begin{tabular}{cccc}
\toprule
\textbf{Method} & \textbf{Problem Linearity} & \textbf{Stochastic Optimization} & \textbf{Citation} \\
\midrule
Neural approximation & Nonlinear & No & \cite{dumouchelle2024neur2bilo} \\
Genetic & Nonlinear & No & \cite{miao2016genetic} \\
Branch and cut & Linear & No & \cite{tahernejad2020branch} \\
Heuristic & Linear & No & \cite{denegre2011interdiction} \\
Heuristic & Nonlinear & No & \cite{soares2023deterministic} \\
Exhaustive enumeration & Linear, Nonlinear & No & \cite{tang2016class} \\
Multi-parametric programming & Linear, Quadratic & No & \cite{avraamidou2017mixed} \\
\bottomrule
\end{tabular}
}
\label{tab:MIBO methods}
\end{table*}

\section{Task Fusion Model}
\begin{figure}[htb]
\centering
\includegraphics[width=\columnwidth,trim=0 60 0 0, clip]{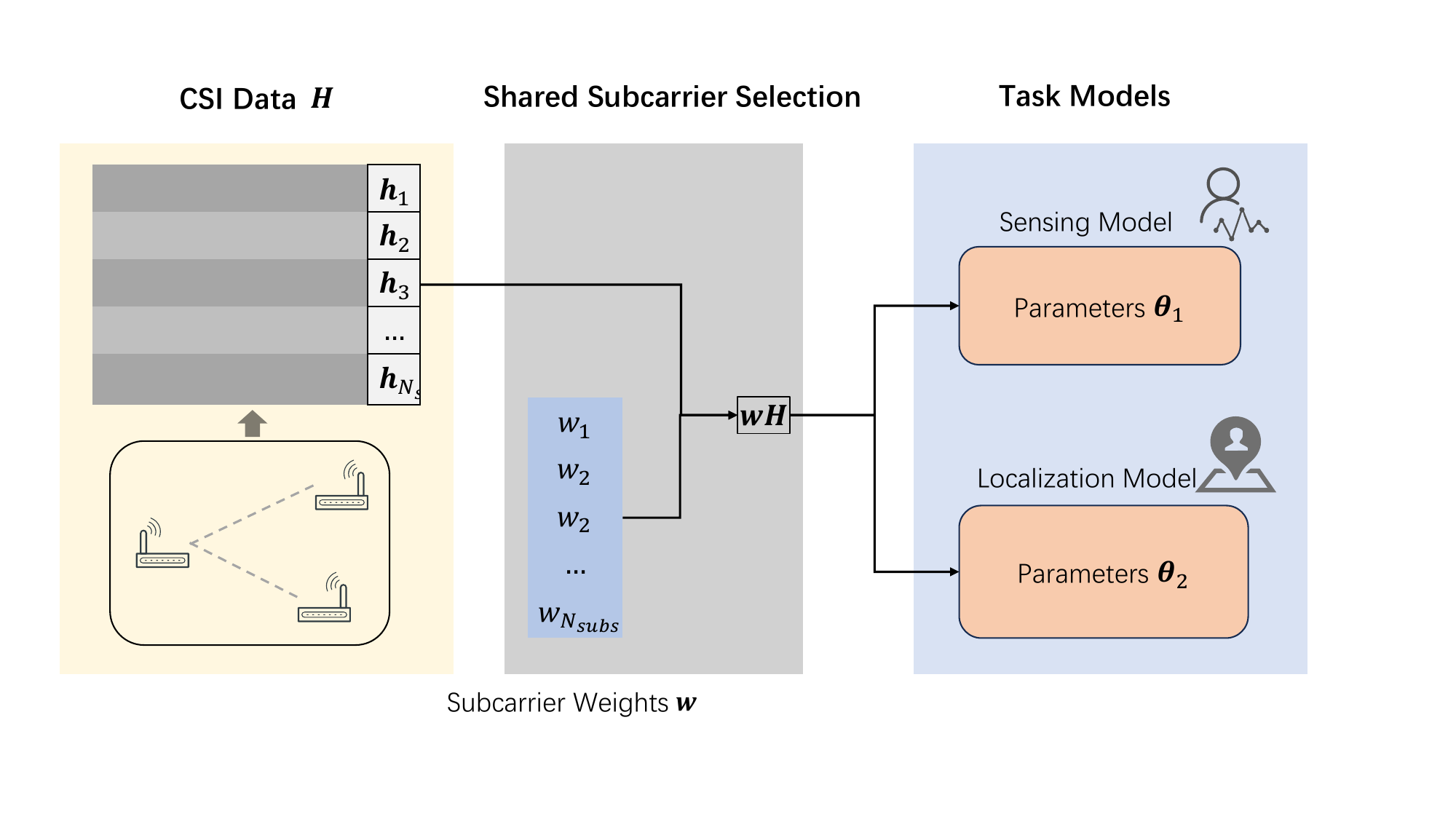} 
\caption{Shared Subcarrier Selection for Task Models Learning}
\label{shared learning}
\end{figure}

This section introduces the formulation of a joint learning framework for passive wireless localization and sensing. We first describe the typical system setup and task structure, and then motivate our model fusion design based on the hierarchical relationship between localization and sensing. Finally, we provide a bilevel optimization formulation for training the proposed system (see Fig. \ref{shared learning}).

\subsection{System Setup and Learning Objectives}
A typical passive wireless localization and sensing system consists of $N_{TX}$ transmitters and $N_{RX}$ receivers deployed in an indoor environment, where $N_{\text{TX}}$ and $N_{\text{RX}}$ denote the number of transmitters and receivers, respectively, as illustrated in Fig. \ref{Joint scenario}. Following prior works on fingerprint localization \cite{yiu2015gaussian,chen2016robust} and device-free sensing \cite{wu2022wifi}, A human target in the scene is characterized by its spatial location $\bm{p}$ and a physical state measurement $\bm{m}$ (e.g., motion, activity, respiration). The wireless propagation channel is affected by the target, and the receivers collect the resulting Channel State Information $\bm{H}$ as high-dimensional input. Machine learning-based approaches typically employ deep neural networks with large-scale parameters $\bm{\theta}$ to extract relevant features from $\bm{H}$ for predicting both $\bm{p}$ and $\bm{m}$. 

Let $\bm{H} \in \mathbb{C}^{N_{pair} \times M}$ denote the input CSI matrix, where $N_{pair}=N_{TX}\times N_{RX}$ is the number of transmitter and receiver pairs, $M$ is the number of subcarriers of each node, and $N_{subs}=N_{pair}\times M$ refers to the total number of subcarriers in the system. Let $\bm{w} \in \{0,1\}^{N_{subs}}$ represent the binary subcarrier selection vector. We define two deep neural network models as $\mathcal{F}_s(\bm{\theta}_1, \bm{w}, \bm{H})$, $\mathcal{F}_l(\bm{\theta}_2, \bm{w}, \bm{H})$ for sensing and localization, respectively. The shared subcarrier selection vector $\bm{w}$ is treated as the parameter of a shared selection module applied to both tasks.

Given the same input $\bm{H}$, the two models produce predictions $\hat{\bm{m}}$ and $\hat{\bm{p}}$, respectively. We employ task-specific loss functions $\mathcal{L}_s(\bm{m}, \hat{\bm{m}})$ and $\mathcal{L}_l(\bm{p}, \hat{\bm{p}})$ for sensing and localization, respectively. The objectives are to minimize both the sensing model loss $\mathcal{L}_s$ and the localization model loss $\mathcal{L}_l$ by optimizing the task-specific parameters $\bm{\theta}_1$ and $\bm{\theta}_2$, while jointly optimizing the shared subcarrier selection vector $\bm{w}$.

Thus, we present a fusion design of wireless localization and sensing with shared subcarrier selection as shown in Fig. \ref{shared learning}, where the parameters of subcarrier selection module are represented as $\bm{w}$ with $\{0,1\}$ integer constraints. $\bm{H}$ is the CSI matrix obtained from a MIMO-OFDM system.

\subsection{MIBO Formulation of Synergistic Training}
Since bilevel optimization is widely used in multi-objective machine learning optimization problems with hierarchical structures and there exists $\{0,1\}$ integer-constrained variables in our models, we precent a MIBO formulation of our localization and sensing model synergistic fusion training problem as follows:
\begin{equation}
\begin{aligned}
    \label{bilevel}
    &\min_{\substack{\bm{\theta}_2} }  \mathcal{L}_s( \{\bm{w}_{i}^*\},\bm{\theta}_2,\{\bm{H}_i\},\{\bm{m}_i\})\\
    &s.t.\quad  \bm{w}_{i}^* \in \{0,1\}^{N_{subs}},i=1,...,N\\
    &\qquad \   N_{min}\leq\|\bm{w}_i^*\|_1\leq N_{max}\\
    &\qquad \  \{\bm{w}_{i}^*\}= \arg\min_{\substack{\{\bm{w}_i\},\bm{\theta}_1}}\ \mathcal{L}_l(\{\bm{w}_{i}\},\bm{\theta}_1,\{\bm{H}_i\},\{\bm{p}_i\}) \\
    & \qquad \  (\bm{H_i},\bm{m}_i,\bm{p}_i) \in \mathbb{D} \\
    & var \quad \bm{\theta}_1,\bm{\theta}_2,\{\bm{w}_i\},
\end{aligned} 
\end{equation}
where $\mathcal{L}_s,\mathcal{L}_l$ refers to the loss of the sensing model and the localization model, respectively. And $\bm{\theta}_1,\bm{\theta}_2$ are the parameters of the localization model and the sensing model, respectively. The use of $\{\bm{H}_i\},\{\bm{m}_i\}, \{\bm{p}_i\}$ in the loss functions reflects that these functions operates on a batch of randomly sampled data instances. $ N_{min}, N_{max}$  represent the minimum and maximum numbers of selected subcarriers, respectively, which are two hyperparameters used to control the subcarrier selection. $\mathbb{D}=\{(\bm{H}_i,\bm{m}_i,\bm{p}_i)\},i=1,...,N$ is the dataset with $N$ samples. $\bm{H}_i,\bm{m}_i,\bm{p}_i$ represents the CSI matrix, the sensing measurement time series (or state vectors) and the position vector of the $i_{th}$ sample, respectively. $\bm{w}_{i}$ constitutes the binary weights to be optimized of the subcarriers in the $i_{th}$ data sample. $\|\bm{w}_i^*\|_1$ are the $l_1$ norm of the the lower-level optimized weights vector. $\bm{w}_i$ represents for subcarrier weights of the $i_{th}$ sample, which is shared for the lower-level problem and the upper-level problem. Since the sensing task requires more granular CSI data features compared to the localization task, the sensing model training loss function is set as the upper-level optimization objective, which utilizes the optimal weights that minimize the lower-level localization model training loss function. 

\section{Stochastic Proximal Mixed-Integer Bilevel Optimization Algorithm}
In this section, we precent the SPG-MIBO algorithm to solve the optimization problem (\ref{bilevel}).  

Solving problem (\ref{bilevel}) is quite challenging due to the presence of $\{0,1\}$ integer variables in the upper and lower-level objective functions, which makes the functions non-continuous and non-smooth. Additionally, the optimization functions in both upper and lower-levels are non-convex functions with large-scale parameters. Thus, we mainly take three steps to designed the algorithm: 1) We first give a relaxation for the lower-level optimal constraint and utilize the Lagrangian dual function to get a single level approximation formulation. 2) The $\{0,1\}$ integer constraint is relaxed and a differentiable mapping function is designed to tighten the relaxation. 3) An SPG method with cutting plane updating is designed to solve the MIBO problem.

\subsection{MIBO Problem Relaxation and Tightening Scheme}
\subsubsection{lower-level Approximation}
The lower-level optimal constraint can be approximated as follows.
\begin{equation}
    \label{lower relaxation}
    \|\mathcal{L}_l(\{\bm{w}_i\},\bm{\theta}_1,\{\bm{H}_i\},\{\bm{p}_i\})- \mathcal{L}_l(\{\bm{w}_i^*\},\bm{\theta}_1^*,\{\bm{H}_i\},\{\bm{p}_i\})\|^2\leq \epsilon,
\end{equation}
where $\mathcal{L}_l(\{\bm{w}_i^*\},\bm{\theta}_1^*,\{\bm{H}_i\},\{\bm{p}_i\})$ refers to the loss with optimal solution $\bm{\theta}_1^*, \{\bm{w}_i^*\}$. In practice, solving $\bm{\theta}_1^*,\{\bm{w}_i^*\}$ is approximated with $K$ steps gradient descent solution $\hat{\bm{\theta}_1},\{\hat{\bm{w}_i}\}$ of lower-level problem. Define
 \begin{equation}
 \begin{aligned}
    &\mathcal{J}(\{\bm{w}_i\},\bm{\theta}_1,\{\bm{H}_i\},\{\bm{p}_i\})\\
    = &\| \mathcal{L}_l(\{\bm{w}_i\},\bm{\theta}_1,\{\bm{H}_i\},\{\bm{p}_i\})- \mathcal{L}_l(\{\hat{\bm{w}_i}\},\hat{\bm{\theta}_1},\{\bm{H}_i\},\{\bm{p}_i\})\|^2.
 \end{aligned}
 \end{equation}
Following the cutting plane method proposed in \cite{yang2008distributed,yang2014distributed,jiao2022asynchronous}, the feasible region of lower-level variables $\{\bm{w}_i\},\bm{\theta}_1$ with constraint $\mathcal{J}(\{\bm{w}_i\},\bm{\theta}_1,\{\bm{H}_i\},\{\bm{p}_i\})\leq \epsilon$ can be further approximated by a polytope as follows.
 \begin{equation}
\label{ploytope}
    \mathbb{P}^t = \{\sum_{i=1}^{N_{subs}} \bm{a}_{i,l} \bm{w}_i + \bm{b}_l\bm{\theta}_1 +\bm{c}_l \leq 0\},l=1,...,|\mathbb{P}^t|,
\end{equation}
where $\mathbb{P}^t$ is the approximation polytope in the $t_{th}$ iteration with $|\mathbb{P}^t|$ cutting planes, $ \bm{a}_{i,l}, \bm{b}_l, \bm{c}_l$ are the coefficents of the $l_{th}$ cutting plane. 

This approach offers the advantage of relaxing the inherently non-convex feasible region arising from the optimality conditions of the lower-level problem into a tractable convex set. Moreover, it gives a possibility to progressively tightens the relaxation by iteratively refining this relaxed feasible region. In addition, this approach allows the computation of different cutting-plane constraints in an asynchronous manner \cite{yang2008distributed,yang2014distribut} thereby facilitating parallel acceleration of optimization in high-dimensional machine leaning.
\subsubsection{Tighten lower-level Approximation}
Also, the approximation of the lower-level optimal constraint with $\mathbb{P}^t$ can be tightened by cutting planes updating with:
\begin{equation}
\label{cutting plane update}
     \mathbb{P}^{t+1}=\begin{cases}
   Drop(\mathbb{P}^{t},cp_i^l) & \text{if } \nu_i^l =0 \  and \ \nu_i^{l+1} =0, \\
   Add(\mathbb{P}^{t},cp_i^{new}) & \text{if } \mathcal{J}(\bm{w}_i^{t+1},\bm{\theta}_1^{t+1},\bm{H}_i,\bm{p}_i)>\epsilon, \\
    \mathbb{P}^{t} & \text{else} 
\end{cases}
\end{equation}
where $cp_i^l$ represents the $l_{th}$ cutting plane for the feasible region constraint. Next, new cutting planes can be added in the $t_{th}$ iteration as follows.
\begin{equation}
\begin{aligned}
    \mathcal{J}(\{\bm{w}_i^{t+1}\},\bm{\theta}_1^{t+1},\{\bm{H}_i\},\{\bm{p}_i\})+&\\
\left[
\begin{array}{cc}
   \frac{\partial\mathcal{J}(\{\bm{w}_i^{t+1}\},\bm{\theta}_1^{t+1},\{\bm{H}_i\},\{\bm{p}_i\} }{\partial\bm{w}_i^{t+1}}\\
     \frac{\partial\mathcal{J}(\{\bm{w}_i^{t+1}\},\bm{\theta}_1^{t+1},\{\bm{H}_i\},\{\bm{p}_i\} }{\partial\bm{\theta}_1^{t+1}}
\end{array}
\right]^T
\left[
\begin{array}{cc}
         \{\bm{w}_i\} -\{\bm{w}_i^{t+1}\} \\
        \bm{\theta}_1-\bm{\theta}_1^{t+1}
\end{array}
\right] &\leq \epsilon.
\end{aligned}
\end{equation}
\subsubsection{Relaxation of Integer Constraint}
We first relax the $\{0,1\}^d$ space constrained $\bm{w}$ to $[0,1]^d$ space as $\bar{\bm{w}}$, then introduce a regularization term $\mathcal{G}(\bar{\bm{w}})$ to represent the distance between $\bar{\bm{w}}$ and the constraint space, where $\mathcal{G}(\bar{\bm{w}})$ is defined as follows.
\begin{equation}
\label{regularization function1}
\begin{aligned}
     &\mathcal{G}(\bar{\bm{w}}) = \|\bar{\bm{w}}-\hat{\bm{x}}\|^2\\
     &s.t.\quad  \hat{\bm{x}}=\arg\min_{\bm{x}} \|\bar{\bm{w}}-\bar{\bm{x}}\|^2, \\
     &\qquad \  \bm{x}\in\{0,1\}^d,\bar{\bm{w}}\in [0,1]^d.
\end{aligned}
\end{equation}
\subsection{Lagrangian Relaxation and Reformulation}
With the $\{0,1\}$ integer constraint and relaxation of the lower-level optimality constraint, the MIBO problem (\ref{bilevel}) can be transformed as follows.
\begin{equation}
\begin{aligned}
    \label{lower-level and integer approximation}
    &\min_{\substack{\bm{\theta}_2} }  \mathcal{L}_s( \{\bar{\bm{w}}_{i}\},\bm{\theta}_2,\{\bm{H}_i\},\{\bm{m}_i\})+\mathcal{G}(\{\bar{\bm{w}}_{i}\})\\
    & s.t.\quad  N_{min}\leq\|\bar{\bm{w}}_i\|_1\leq N_{max},\quad i=1,...,N\\
    & \qquad \sum_{i=1}^{N} \bm{a}_{i,l} \bar{\bm{w}}_i + \bm{b}_l\bm{\theta}_1 +\bm{c}_l \leq 0,\quad l=1,...,|\mathbb{P}^t|\\
    & \qquad  (\bm{H_i},\bm{m}_i,\bm{p}_i) \in \mathbb{D} \\
    & var \quad \bm{\theta}_1,\bm{\theta}_2,\{\bar{\bm{w}}_i\}.
\end{aligned}
\end{equation}

Since the transformed problem is a single-level optimization problem only with inequality constraints, Lagrangian relaxation can be applied to solve the problem; thus, we have the following Lagrangian function.
\begin{equation}
\label{lagrangian function}
\begin{aligned}
     &\mathcal{L}( \{\bar{\bm{w}}_{i}\},\bm{\theta}_1,\bm{\theta}_2,\{\bm{H}_i\},\{\bm{m}_i\},\bm{\lambda},\bm{\mu})+\mathcal{G}(\{\bar{\bm{w}}_{i}\})\\  
      =& \mathcal{L}_s( \{\bar{\bm{w}}_{i}\},\bm{\theta}_2,\{\bm{H}_i\},\{\bm{m}_i\})\\
      &+\sum_{l=1}^{|\mathbb{P}|}\lambda_l \mathcal{C}_l+\sum_{i=1}^N\bm{\mu}_i\mathcal{S}_i+\mathcal{G}(\{\bar{\bm{w}}_{i}\}),
\end{aligned}
\end{equation}
where $\sum_{l=1}^{|\mathbb{P}|}\lambda_l \mathcal{C}_l$ and $ \sum_{i=1}^N\bm{\mu}_i\mathcal{R}_i$ are the Lagrange multipliers terms with Lagrange multipliers $\lambda_l$ and $\mu_i$. $\lambda_l\mathcal{C}_l$ and $\bm{\mu_i}\mathcal{S}_i$ are defined as follows.
\begin{equation}
    \begin{aligned}
        \lambda_l\mathcal{C}_l&=\lambda_l(\sum_{i=1}^{N} \bm{a}_{i,l} \bar{\bm{w}}_i + \bm{b}_l\bm{\theta}_1 +\bm{c}_l),\\
        \bm{\mu}_i\mathcal{S}_i&= \mu_{1,i}(N_{min}-\|\bar{\bm{w}_i}\|_1) \\
        &+\mu_{2,i} (\|\bar{\bm{w}_i}\|_1-N_{max}).
    \end{aligned}
\end{equation}
\subsection{Stochastic Proximal Gradient Method}
With Lagrangian function (\ref{lagrangian function}), the variables updated with the Proximal Gradient Descent method for the relaxed variables and Gradient Descent, Reverse Gradient Descent for continuous variable and Lagrange multipliers, are respectively as follows.
\begin{equation}
\label{update}
\begin{aligned}
\begin{cases}
\tilde{\bm{w}_i}^t=\bar{\bm{w}_i}^{t}-\eta_t\nabla_{\bar{\bm{w}_i}^t}\mathcal{L}( \{\bar{\bm{w}_{i}}\},\bm{\theta}_2,\{\bm{H}_i\},\{\bm{m}_i\},\bm{\lambda},\bm{\mu})\\ 
\bar{\bm{w}_i}^{t+1}= \text{prox}_{\eta_t,\mathcal{G}}(\tilde{\bm{w}_i}^t)\\
\bm{\theta}_1^{t+1}= \bm{\theta}_1^{t}- \eta_t\nabla_{\bm{\theta}_1^t}\mathcal{L}( \{\bar{\bm{w}_{i}}\},\bm{\theta}_2,\{\bm{H}_i\},\{\bm{m}_i\},\bm{\lambda},\bm{\mu})\\
\bm{\theta}_2^{t+1}=\bm{\theta}_2^{t}- \eta_t\nabla_{\bm{\theta}_2^t}\mathcal{L}( \{\bar{\bm{w}_{i}}\},\bm{\theta}_2,\{\bm{H}_i\},\{\bm{m}_i\},\bm{\lambda},\bm{\mu})\\
\bm{\lambda}^{t+1}=\bm{\lambda}^{t}+\eta_t\nabla_{\bm{\lambda}^t}\mathcal{L}( \{\bar{\bm{w}_{i}}\},\bm{\theta}_2,\{\bm{H}_i\},\{\bm{m}_i\},\bm{\lambda},\bm{\mu})\\
\bm{\mu}^{t+1}=\bm{\mu}^{t}+\eta_t\nabla_{\bm{\mu}^t}\mathcal{L}( \{\bar{\bm{w}_{i}}\},\bm{\theta}_2,\{\bm{H}_i\},\{\bm{m}_i\},\bm{\lambda},\bm{\mu}),
\end{cases}
\end{aligned}
\end{equation}
where the proximal operation is defined as
\begin{equation}
\label{proximal gradient func}
    prox_{\eta_t,\mathcal{G}}(\bar{\bm{w}}) = \arg\min_{\bm{x}} \left(\frac{1}{2\eta_t} \|\bm{x} - \bar{\bm{w}}\|^2 + \mathcal{G}(\bm{x}) \right).
\end{equation}
The above proximal operation is equivalent to the following closed-form expression.
\begin{equation}
    prox_{\eta_t, \mathcal{G}}(\bar{\bm{w}}) = \left[ 
    \begin{cases}
        \dfrac{\bar{w}_i / \eta_t}{\frac{1}{\eta_t} + 2}, & \text{if } \bar{w}_i \in [0, \frac{1}{2}] \\
        \dfrac{\bar{w}_i / \eta_t + 2}{\frac{1}{\eta_t} + 2}, & \text{if } \bar{w}_i \in (\frac{1}{2}, 1]
    \end{cases}
    \right]_{i=1}^d.
\end{equation}
\subsection{Algorithm}
Based on the relaxation and tightening scheme and the derived gradient update rules, the proposed SPG-MIBO algorithm consists of four main steps as follows.
\begin{itemize}
    \item Estimating the optimal solution of the lower-level problem;
    \item Performing stochastic updates for the Lagrangian relaxation problem;
    \item Performing stochastic proximal gradient updates the variables relaxed from the integer constraint; and
    \item Refining the feasible region by updating the polytope with cutting planes.
\end{itemize}
The algorithm ~\ref{alg:spg-mibo} provides a detailed description of the proposed SPG-MIBO method.
\begin{algorithm}[t]
\caption{SPG-MIBO Algorithm}
\label{alg:spg-mibo}
\begin{algorithmic}[1]
\REQUIRE Dataset $\mathbb{D}$, relaxation bound $\epsilon$, initial step-size $\eta$
\ENSURE Optimized parameters $\bm{\theta}_1^*, \bm{\theta}_2^*, \bar{\bm{w}}^*$
\STATE \textbf{Initialize:} $t=0$, $\{\bm{w}_i^0\}$, $\bm{\theta}_1^0$, $\bm{\theta}_2^0$, $\bm{\lambda}^0$, $\bm{\mu}^0$, $\mathbb{P}^0$
\WHILE{not converged}
    \STATE Estimate lower-level optimum: gradient descent for $K$ steps to get $\{\hat{\bm{w}}_i^t\}, \hat{\bm{\theta}}_1^t$
    \STATE Update parameters with stochastic gradient to obtain \\ $\bm{\theta}_1^{t+1}$, $\bm{\theta}_2^{t+1}$, $\bm{\lambda}^{t+1}$, $\bm{\mu}^{t+1}$ \eqref{update}
    \STATE Performing stochastic proximal gradient update to get \\$\bar{\bm{w}}^{t+1}$\eqref{proximal gradient func}
    \STATE Lower-level relaxation tighten \eqref{cutting plane update}\\
    Drop cutting planes and 
        \IF{lower-level constraints violated}
            \STATE Update polytope $\mathbb{P}^{t+1}$ via cutting-plane method
        \ENDIF
\ENDWHILE
\RETURN $\bar{\bm{w}}^*$, $\bm{\theta}_1^*$, $\bm{\theta}_2^*$
\end{algorithmic}
\end{algorithm}
\section{Convergence Analysis of SPG-MIBO}
In this section, we analyze the convergence rate of the proposed SPG-MIBO algorithm. For notational simplicity, we categorize the variables in Equation~(\ref{update}) into three groups: (i) continuous dual variables $\bm{v}$, (ii) other continuous variables $\bm{u}$, and (iii) relaxed variables $\bm{w}$. Specifically, $\bm{u}$ includes $\bm{\theta}_1$ and $\bm{\theta}_2$, $\bm{v}$ includes $\bm{\lambda}$ and $\bm{\mu}$, and $\bm{w}$ denotes $\{\bar{\bm{w}}_i\}$ in this section.

We analyze the convergence rate for each group of variables separately. Since each group achieves a convergence rate of $\mathcal{O}(1/\sqrt{T})$, the overall convergence rate of the proposed SPG-MIBO algorithm is also $\mathcal{O}(1/\sqrt{T})$ under standard assumptions.

Following the work of Ghadimi et al.~\cite{ghadimi2020single} and Jiao et al.~\cite{jiao2022asynchronous}, we adopt the following widely used assumption. 

\textbf{Assumption 1 (L-smoothness)}: The function $\mathcal{L}(\bm{u}, \bm{v}, \bm{w})$ is $L$-smooth, i.e.,
\begin{equation}
\begin{aligned}
   &\mathcal{L}(\bm{u}_{k+1}, \bm{v}_{k+1}, \bm{w}_{k+1})\\
    \leq & \mathcal{L}(\bm{u}_k, \bm{v}_k, \bm{w}_k) 
     + 
    \begin{pmatrix}
        \nabla_{\bm{u}} \mathcal{L}(\bm{u}_k, \bm{v}_k, \bm{w}_k) \\
        \nabla_{\bm{v}} \mathcal{L}(\bm{u}_k, \bm{v}_k, \bm{w}_k) \\
        \nabla_{\bm{w}} \mathcal{L}(\bm{u}_k, \bm{v}_k, \bm{w}_k)
    \end{pmatrix}^T 
    \begin{pmatrix}
        \bm{u}_{k+1} - \bm{u}_k \\
        \bm{v}_{k+1} - \bm{v}_k \\
        \bm{w}_{k+1} - \bm{w}_k
    \end{pmatrix} \\
     &+ \frac{L}{2} \left\| \begin{pmatrix}
        \bm{u}_{k+1} - \bm{u}_k \\
        \bm{v}_{k+1} - \bm{v}_k \\
        \bm{w}_{k+1} - \bm{w}_k
    \end{pmatrix} \right\|^2,
\end{aligned}
\end{equation}
where $\bm{u}\in \mathbb{R}^{n_1}$,$\bm{v}\in \mathbb{R}^{n_2}$,$\bm{w}\in [0,1]^d $ represent continuous variables, dual variables, and relaxed $\{0, 1\}$ constrained variables respectively. The subscript $k$ denotes the $k_{th}$ iteration of the algorithm.

\textbf{Assumption 2 (Stochastic Gradient)}: The stochastic gradient is an unbiased estimate of the gradient, i.e.,
\begin{equation}
    \begin{aligned}
        &\mathbb{E}[\nabla_{\bm{w}} \mathcal{L}(\bm{u}_k, \bm{v}_k, \bm{w}_k; \bm{\xi}_k)]=\nabla_{\bm{w}} \mathcal{L}(\bm{u}_k, \bm{v}_k, \bm{w}_k)\\
        &\mathbb{E}[\nabla_{\bm{u}} \mathcal{L}(\bm{u}_k, \bm{v}_k, \bm{w}_k; \bm{\xi}_k)]=\nabla_{\bm{u}} \mathcal{L}(\bm{u}_k, \bm{v}_k, \bm{w}_k)\\
        &\mathbb{E}[\nabla_{\bm{v}} \mathcal{L}(\bm{u}_k, \bm{v}_k, \bm{w}_k; \bm{\xi}_k)]=\nabla_{\bm{v}} \mathcal{L}(\bm{u}_k, \bm{v}_k, \bm{w}_k).
    \end{aligned}
\end{equation}
\textbf{Assumption 3 (Boundedness)}: The gradients are assumed to have upper bounds, i.e., 
\begin{equation}
    \begin{aligned}
    \|\nabla_{\bm{u}} \mathcal{L}(\bm{u}_k, \bm{v}_k, \bm{w}_k)\|\leq G_{\bm{u}}\\
    \|\nabla_{\bm{v}} \mathcal{L}(\bm{u}_k, \bm{v}_k, \bm{w}_k)\|\leq G_{\bm{v}}\\
    \|\nabla_{\bm{w}} \mathcal{L}(\bm{u}_k, \bm{v}_k, \bm{w}_k)\|\leq G_{\bm{w}}.\\
    \end{aligned}
\end{equation}
The function $\mathcal{F}=\mathcal{L}+\mathcal{G}$ has a lower bound, i.e.,
\begin{equation}
     \mathcal{F}^*\leq \mathcal{F}(\bm{u}, \bm{v}, \bm{w}).
\end{equation}
The variance of the stochastic gradients has upper bounds, i.e.,
\begin{equation}
\begin{aligned}
        &\mathbb{E}[\nabla_{\bm{w}} \mathcal{L}(\bm{u}_k, \bm{v}_k, \bm{w}_k; \bm{\xi}_k)-\nabla_{\bm{w}} \mathcal{L}(\bm{u}_k, \bm{v}_k, \bm{w}_k)]\leq \sigma_{\bm{w}}\\
        &\mathbb{E}[\nabla_{\bm{u}} \mathcal{L}(\bm{u}_k, \bm{v}_k, \bm{w}_k; \bm{\xi}_k)-\nabla_{\bm{u}} \mathcal{L}(\bm{u}_k, \bm{v}_k, \bm{w}_k)]\leq \sigma_{\bm{u}}\\
        &\mathbb{E}[\nabla_{\bm{v}} \mathcal{L}(\bm{u}_k, \bm{v}_k, \bm{w}_k; \bm{\xi}_k)-\nabla_{\bm{v}} \mathcal{L}(\bm{u}_k, \bm{v}_k, \bm{w}_k)]\leq \sigma_{\bm{v}},
\end{aligned}
\end{equation}
where $ \bm{\xi}_k$ are the batch samples in $k_{th}$ step.

\textbf{Theorem 1.}
Under the above assumptions, define the stationary condition of relaxed variables $\bm{w}$ as $\frac{1}{T} \sum_{k=1}^T \mathbb{E}[\|\bm{w}_{k+1} - \bm{w}_k\|^2] $. Let the step size be $\eta_t=\frac{\eta}{\sqrt{t}}$, Then the following holds:
\begin{equation}
\begin{aligned}
    &\frac{1}{T} \sum_{k=1}^T \mathbb{E}[\|\bm{w}_{k+1} - \bm{w}_k\|^2]\\
    \leq & \left( \frac{6\eta}{2\sqrt{T}+\frac{2}{T}-3\eta L}\right)\left(\mathcal{F}(\bm{u}_{1}, \bm{v}_{1}, \bm{w}_{1})-\mathcal{F}^*\right)\\
   &+\left(\frac{L\eta^2\log(T)}{2T} -\frac{2\eta}{\sqrt{T}}\right)\left(G_{\bm{u}}^2+\sigma_{\bm{u}}\right)\\
   &+\left(\frac{L\eta^2\log(T)}{2T} + \frac{2\eta}{\sqrt{T}}\right)\left(G_{\bm{v}}^2+\sigma_{\bm{v}}\right).
\end{aligned}
\end{equation}
Therefore, the convergence rate of $\bm{w}$ is $\mathcal{O}(\frac{1}{\sqrt{T}})$.

\textbf{Theorem 2.} Under the above assumptions, let the stationary condition of the variable $\bm{u}$ be defined as $\frac{1}{T}\sum_{k=1}^T\mathbb{E}[\|\nabla_{\bm{u}} \mathcal{L}(\bm{u}_k, \bm{v}_k, \bm{w}_k)\|^2]$ and taking step size $\eta_t=\frac{\eta}{\sqrt{t}}$. According to Theorem 1, the change of the variable within $T$ consecutive iterations is bounded, i.e., 
$\mathbb{E}[\|\bm{w}_{k+1} - \bm{w}_k\|\leq M_{\bm{w}}$. 
Then we have:
\begin{equation}
    \begin{aligned}
        &\frac{1}{T}\sum_{k=1}^T\mathbb{E}[\|\nabla_{\bm{u}} \mathcal{L}(\bm{u}_k, \bm{v}_k, \bm{w}_k)\|^2]\\
        \leq &  \frac{\mathcal{L}(\bm{u}_1,\bm{v}_1,\bm{w}_1)-\mathcal{L}(\bm{u}_{T+1},\bm{v}_{T+1},\bm{w}_{T+1})}{\mathcal{C}_{\eta,L}(T)}+\frac{L\eta^2\sigma_{\bm{u}}^2\log(T)}{\mathcal{C}_{\eta,L}(T)}\\
        &+\frac{LTM_{\bm{w}}^2}{2\mathcal{C}_{\eta,L}(T)}+\frac{(2\eta\sqrt{T}+\frac{L\eta^2\log(T)}{2})G_{\bm{v}}}{\mathcal{C}_{\eta,L}(T)}+\frac{G_{\bm{w}}TM_{\bm{w}}}{\mathcal{C}_{\eta,L}(T)},
    \end{aligned}
\end{equation}
where $M_{\bm{w}}$ is the upper bound, $\mathcal{C}_{\eta,L}(T) = T(2\eta\sqrt{T}-\frac{L\eta^2\log(T)}{2}))$.

Since $\mathcal{C}_{\eta,L}(T) =\mathcal{O}(T^{1.5})$, it follows that:
\begin{equation}
\begin{aligned}
      \max\{\frac{1}{\mathcal{C}_{\eta,L}(T)},\frac{L\eta^2\sigma_{\bm{u}}^2\log(T)}{\mathcal{C}_{\eta,L}(T)},\frac{LTM_{\bm{w}}^2}{2\mathcal{C}_{\eta,L}(T)},&\\
      \frac{(2\eta\sqrt{T}+\frac{L\eta^2\log(T)}{2})G_{\bm{v}}}{\mathcal{C}_{\eta,L}(T)},\frac{G_{\bm{w}}TM_{\bm{w}}}{\mathcal{C}_{\eta,L}(T)}\}&=\mathcal{O}(\frac{1}{\sqrt{T}}) . 
\end{aligned}
\end{equation}
Thus, the variable $\bm{u}$ achieves a convergence rate of $\mathcal{O}(\frac{1}{\sqrt{T}})$.

\textbf{Theorem 3.} Under the above assumptions, define the stationary condition of variable $\bm{v}$ as $\frac{1}{T}\sum_{k=1}^T\mathbb{E}[\|\nabla_{\bm{v}} \mathcal{L}(\bm{u}_k, \bm{v}_k, \bm{w}_k)\|^2]$ and take step size $\eta_t=\frac{\eta}{\sqrt{t}}$. Then, the following holds:
\begin{equation}
    \begin{aligned}
        &\frac{1}{T}\sum_{k=1}^T\mathbb{E}[\|\nabla_{\bm{v}} \mathcal{L}(\bm{u}_k, \bm{v}_k, \bm{w}_k)\|^2]\\
        \leq &  \frac{\mathcal{L}(\bm{u}_1,\bm{v}_1,\bm{w}_1)-\mathcal{L}(\bm{u}_{T+1},\bm{v}_{T+1},\bm{w}_{T+1})}{\mathcal{C}'_{\eta,L}(T)}+\frac{L\eta^2\sigma_{\bm{v}}^2\log(T)}{\mathcal{C}'_{\eta,L}(T)}\\
        &+\frac{LTM_{\bm{w}}^2}{2\mathcal{C}'_{\eta,L}(T)}+\frac{(\frac{L\eta^2\log(T)}{2}-2\eta\sqrt{T})G_{\bm{u}}}{\mathcal{C}'_{\eta,L}(T)}+\frac{G_{\bm{w}}TM_{\bm{w}}}{\mathcal{C}'_{\eta,L}(T)},
    \end{aligned}
\end{equation}
where $M_{\bm{w}}$ is an upper bound of $\mathbb{E}[\|\bm{w}_{k+1} - \bm{w}_k\|] $ over $T$ steps, and $\mathcal{C}'_{\eta,L}(T) = T(2\eta\sqrt{T}+\frac{L\eta^2\log(T)}{2})$. Since $\mathcal{C}'_{\eta,L}(T) =\mathcal{O}(T^{1.5})$, the max term on the right-hand side satisfies the rate $\mathcal{O}(\frac{1}{\sqrt{T}})$ i.e.,
\begin{equation}
\begin{aligned}
    \max\{\frac{1}{\mathcal{C}'_{\eta,L}(T)},\frac{L\eta^2\sigma_{\bm{v}}^2\log(T)}{\mathcal{C}'_{\eta,L}(T)},\frac{LTM_{\bm{w}}^2}{2\mathcal{C}'_{\eta,L}(T)},&\\
    \frac{(\frac{L\eta^2\log(T)}{2}-2\eta\sqrt{T})G_{\bm{u}}}{\mathcal{C}'_{\eta,L}(T)},\frac{G_{\bm{w}}TM_{\bm{w}}}{\mathcal{C}'_{\eta,L}(T)}\}&=\mathcal{O}(\frac{1}{\sqrt{T}}).
\end{aligned}
\end{equation}
Therefore, $\bm{v}$ has a convergence rate of $\mathcal{O}(\frac{1}{\sqrt{T}})$

All corresponding proofs for Theorems 1–3 are provided in Appendix B.

\section{Experiment}
As mentioned in this paper, our task is to achieve synergistic training of the localization and sensing models. In this section, we take the estimation of human respiration as a specific sensing task. We first briefly introduce the experimental setup and datasets. Then, we compare and analyze the model performance under SPG-MIBO synergistic training versus separately training. Furthermore, we compare the optimization convergence performance and variable optimization results of the proposed proximal projection method with the general penalty function method under integer constraint conditions. Finally, to demonstrate the model-agnostic property of our approach, we evaluated it by comparing with benchmark models across multiple datasets.
\subsection{Experimental Setting and Dataset}

\begin{figure}[t]
\centering
\includegraphics[width=\columnwidth,trim=150 70 120 100, clip]{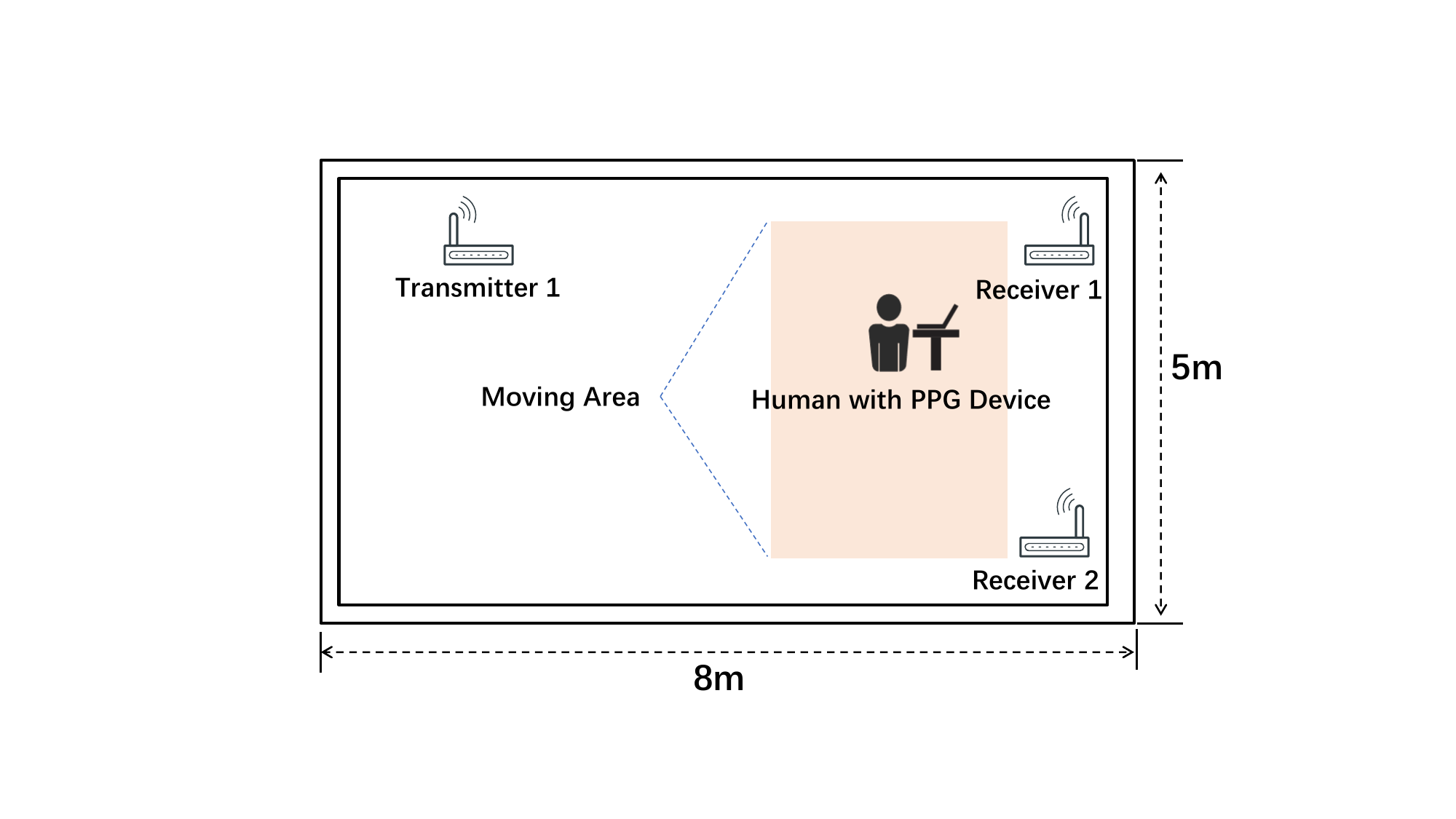} 
\caption{Experimental Setup for Wi-Respiration Dataset Collection}
\label{wi-respiration setup}
\end{figure}

\begin{figure}[t]
\centering
\includegraphics[width=0.7\columnwidth,trim=0 0 0 0, clip]{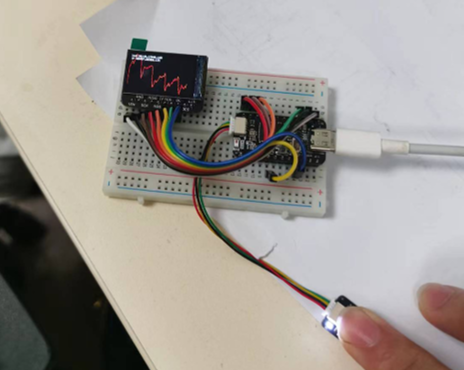} 
\caption{Respiration Pattern Acquisition Using a PPG Biophysical Signal Recorder}
\label{ppg device}
\end{figure}

\begin{figure}[t]
\centering
\includegraphics[width=0.7\columnwidth,trim=0 0 100 100, clip]{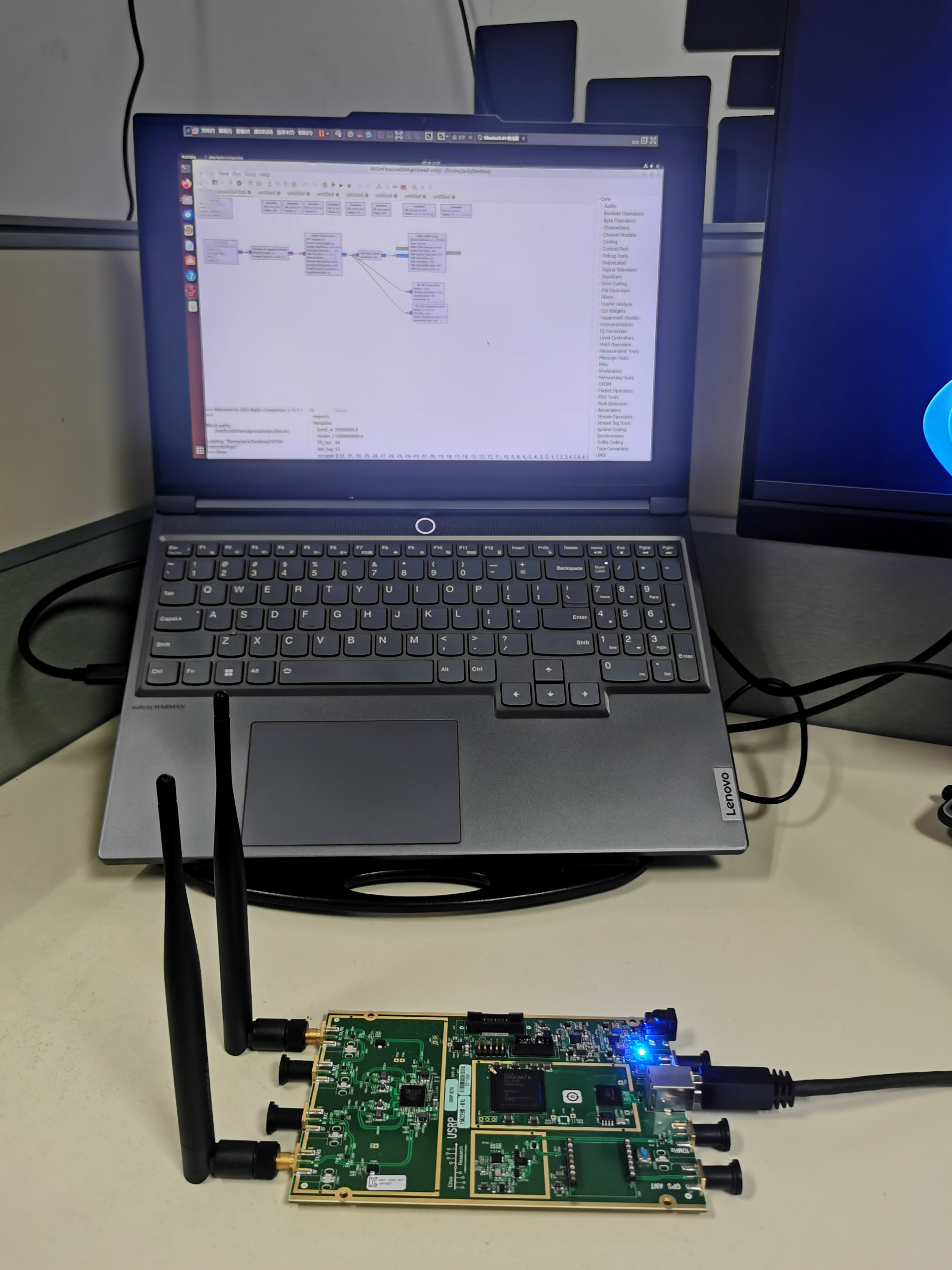} 
\caption{USRP B210-Based Transmitter and Receiver Nodes with GNU Radio Interface}
\label{usrp}
\end{figure}

We utilized three distinct datasets to evaluate the proposed method across various scenarios. 
\begin{itemize}
    \item \textbf{Wi-Respiration Dataset:} This dataset is collected in a typical office environment (see Fig. \ref{wi-respiration setup}) using three USRP B210 (see Fig. \ref{usrp}) devices configured as one TX and two RXs, with each RX acquiring OFDM CSI data $\bm{H}$ comprising 32 subcarriers, resulting in $N_{\text{subs}} = 64$. The sensing target is equipped with a photoplethysmography (PPG) device (see Fig. \ref{ppg device}) to record biophysical respiration signals $\bm{m}$, while the target’s position is annotated as a two-dimensional vector $\bm{p} = [x, y]$ in a Cartesian coordinate system. All measurements $\bm{H}$, $\bm{m}$, and $\bm{p}$ are collected as time series and segmented into short time slots $\Delta t$, during which it is assumed that the target remain stationary. Thus, the dataset is represented as $\mathbb{D} = \{ (\bm{H}_i, \bm{m}_i, \bm{p}_i) \}$, where $i$ denotes the index of each time slot. The task associated with this dataset is fine-grained joint localization and respiration sensing.

    \item \textbf{Simulated Dataset:} To address the challenge of limited data in real-world scenarios, we construct a synthetic dataset based on a classical wireless channel model, detailed in Appendix \ref{appendix:simulation_experiment}. This dataset simulates coarse-grained localization (at the room level) and basic activity recognition tasks, including standing, sitting, and lying down. It provides a controlled environment for large-scale algorithm validation.

    \item \textbf{Wi-Mans Dataset}\cite{huang2024wimans}: This is a publicly available, cross-scene, multimodal dataset containing synchronized Wi-Fi CSI and video data. Supports multiple tasks, including fine-grained activity recognition and the localization of multiple targets with unknown classes. The dataset is collected in diverse indoor real-world settings, making it ideal for evaluating model generalization.
\end{itemize}

In terms of model design, given the inherent challenges of respiration sensing and fine-grained localization in confined environments, we employ recent state-of-the-art architectures for both sensing and localization to ensure robust performance on the Wi-Respiration dataset. Specifically, we refer to recent advances in deep learning-based indoor Wi-Fi CSI localization \cite{rao2023mffaloc} and sensing \cite{he2023robust}. To constrain the parameters of the subcarrier selection module within a valid range, we implement the subcarrier weighting layer using a randomly initialized learnable parameter layer followed by a sigmoid activation. Furthermore, to evaluate the model-agnostic property of the proposed joint optimization algorithm, we adopt the benchmark localization and sensing models from \cite{huang2024wimans} and conduct experiments on both a simulated dataset and the Wi-Mans dataset.

In terms of model training, we employ the commonly used approach in deep learning of partitioning the data into training and validation sets. Additionally, we utilize K-fold cross-validation to robustly evaluate the model’s average performance on the validation sets. Based on experimental observations from the work \cite{yang2022efficientfi}, we set the hyperparameters \( N_{min} \) and \( N_{max} \), which constrain the number of selected subcarriers, to approximately one-third and one-half of \( N_{subs} \) as the lower and upper bounds, respectively.

\subsection{Numerical Results}
\subsubsection{Performance Comparison}
To evaluate the effectiveness of our synergistic fusion strategy that jointly trains the localization and sensing models using SPG-MIBO, we conduct experiments on our real-world dataset Wi-Respiration. For baseline comparisons, the models are also trained separately using standard Stochastic Gradient Descent (SGD), and their performance is measured in terms of Mean Squared Error (MSE). During separate training, we enforce the $\{0,1\}$ constraints on integer-constrained variables using a mainstream quadratic penalty function as defined in \eqref{quaraic_penalty}, where $\lambda_p > 0$ is a penalty coefficient:
\begin{equation}
\label{quaraic_penalty}
    \text{Penalty}(\bm{w}) = \lambda_p \sum_i (1 - w_i) w_i.
\end{equation}

\begin{figure}[htb]
\centering
\includegraphics[width=0.9\columnwidth]{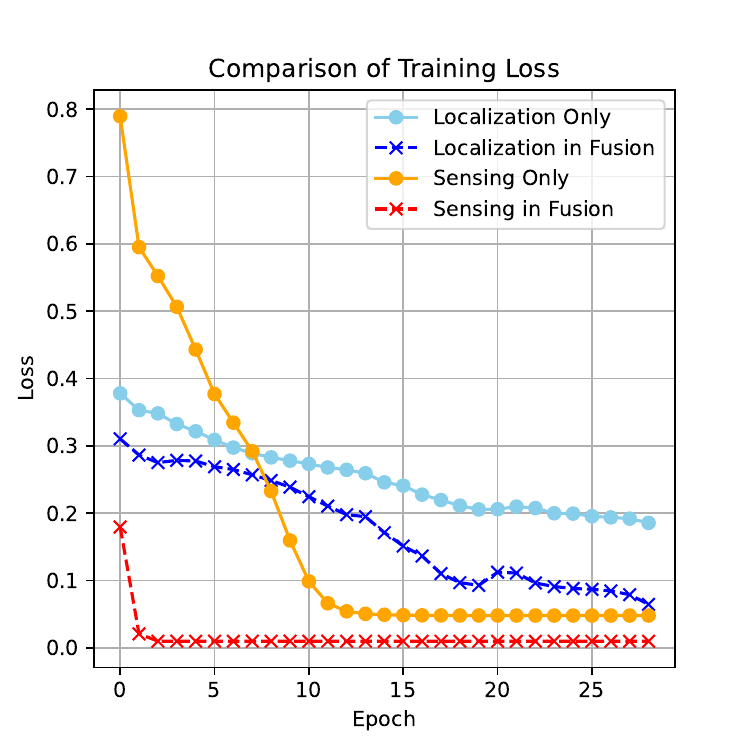} 
\caption{Training Loss Comparison Between Synergistic Fusion Training and Single-Task Training for Localization and Sensing Models Under the Same Model Architecture.}
\label{compare}
\end{figure}

Experimental results, as shown in Fig. \ref{compare}, demonstrate that our proposed SPG-MIBO-based joint training outperforms separate training for both localization and sensing tasks, under the same model architecture. (Note that in Fig. \ref{compare} the model loss at the first epoch under the fusion training mode is significantly lower than that of separate training. This is because the bilevel optimization performs multiple inner-loop iterations (i.e., \(K\) steps) to estimate the lower-level optimal solution before updating the upper-level parameters.)

\subsubsection{Convergence Comparison with SGD}
\begin{figure}[htb]
\centering
\includegraphics[width=0.9\columnwidth]{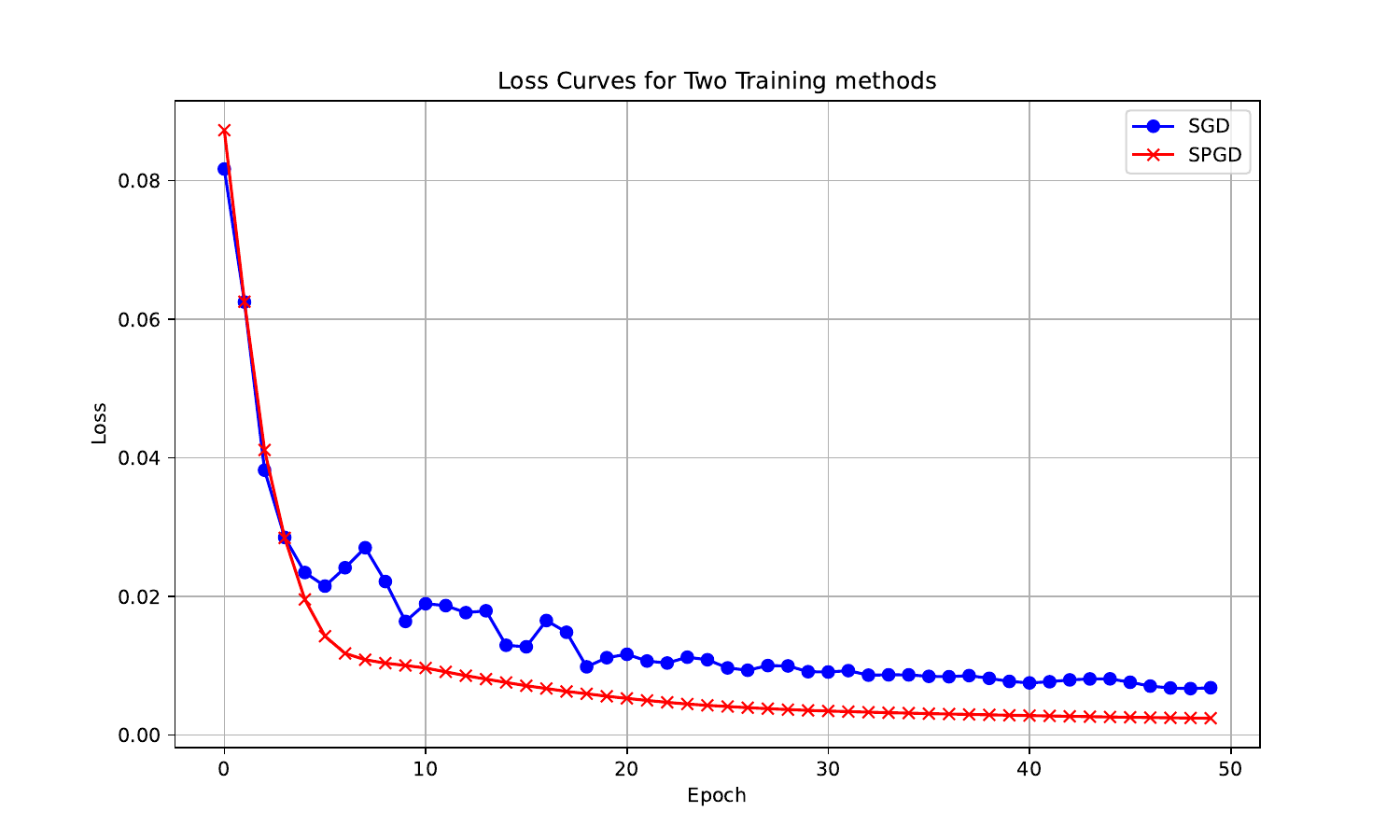} 
\caption{Loss Curves for SGD-Penalty and SPGD with Mixed-Integer Machine Learning Problem for Convergence Rate Comparison}
\label{fig:convergence}
\end{figure}

Our experimental results further demonstrate that the proximal operation introduced in our SPG-MIBO algorithm does not hinder convergence. As shown in Fig.~\ref{fig:convergence}, the loss curves for solving the relaxed mixed-integer learning problem using both the SGD-penalty method and the proposed Stochastic Proximal Gradient Descent (SPGD) method exhibit a similar decreasing trend. Since it is well established that SGD achieves a \( \mathcal{O}(1/\sqrt{T}) \) convergence rate in nonconvex optimization, the comparable behavior observed here suggests that our use of proximal updates preserves the theoretical convergence guarantees.

\subsubsection{Integer-Constrained Variable Analysis}
\begin{figure}[htb]
\centering
\includegraphics[width=0.9\columnwidth]{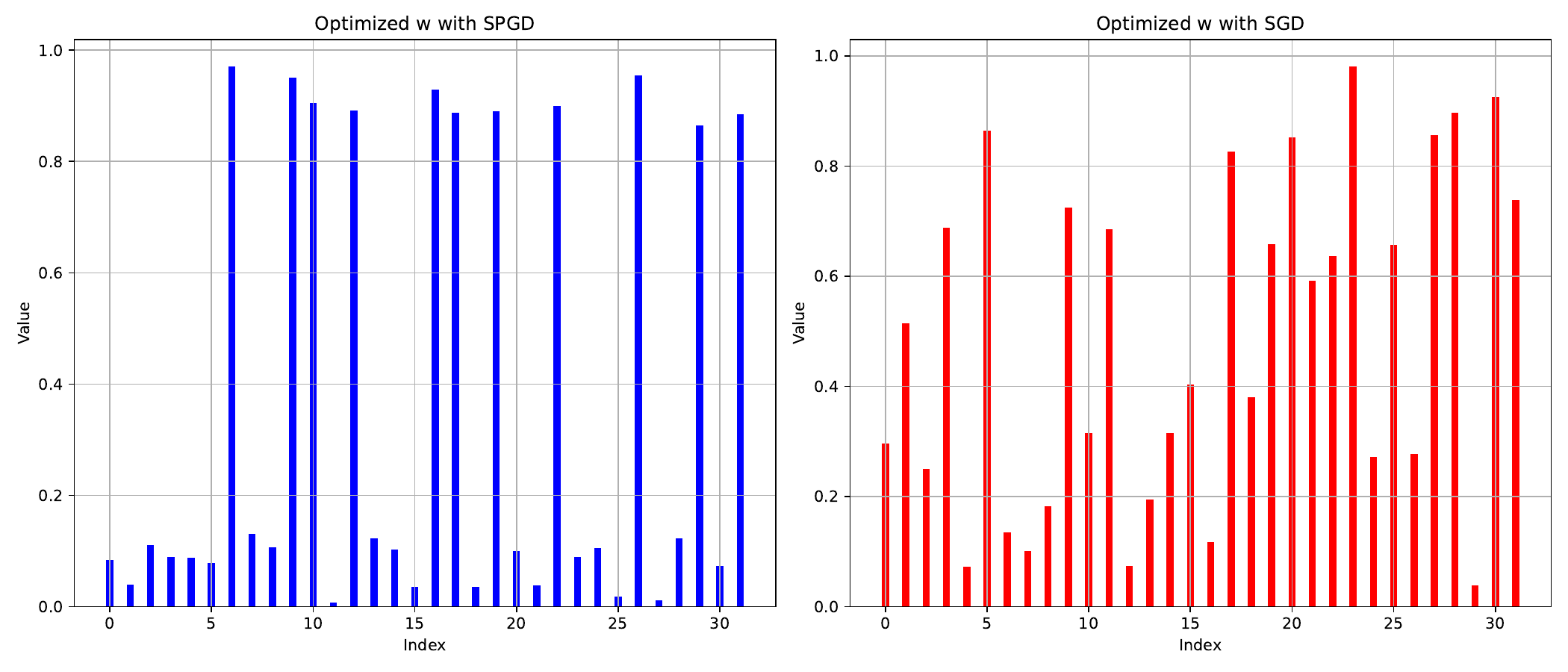} 
\caption{Optimized Relaxed Variables from Integer Constraint with SPGD and SGD-Penalty.}
\label{fig:optimized w}
\end{figure}

In the experiment on the simulation dataset, we selected the signal received from a single RX and visualized the optimized representative subcarrier selection weight vector of the 32 subcarriers. In Fig. \ref{fig:optimized w} the relaxed variables optimized with SPGD and our proximal operation design are much closer to the original $\{0,1\}^d$ space, indicating a tighter relaxation for the $\{0,1\}$ integer constraint. 

We use the Integer Feasibility Gap metric to measure the gap between the integer constrained parameters of the optimized solution and the original integer constrain space. The degree of the integer constraints being satisfied is compared for different methods.

\begin{equation}
\text{Integer Feasibility Gap} = \frac{|x_{\text{relaxed}} - x_{\text{int}}|}{|x_{\text{int}}|}
\end{equation}
where \( x_{\text{relaxed}} \) is the relaxed solution and \( x_{\text{int}} \) is the closest integer solution in the original constrained space.
Table \ref{metric_compare} shows that the SPGD achieves a better satisfication of the integer constraint.

\begin{table}[ht]
\caption{Metric Values for Different Methods}
\centering
\begin{tabular}{lc}
\toprule
\textbf{Method} & \textbf{Metric Value} \\
\midrule
SPG-MIBO & 0.041 \\
Penalty-MIBO & 0.116 \\
\bottomrule
\end{tabular}
\label{metric_compare}
\end{table}

\subsubsection{Generalization and Model-Agnostic Evaluation}

To further validate the generalizability of our modeling approach and algorithm, we conduct experiments on multiple datasets collected from different MIMO-OFDM wireless network scenarios. These scenarios vary in terms of the sensing objectives and localization accuracy requirements. In addition, we employ benchmark model architectures to demonstrate that the performance improvement achieved by our method is model-agnostic and not depend on specific network structures.

Specifically, we employ the same model architecture in benchmark\cite{huang2024wimans} across three MIMO-OFDM Wi-Fi CSI datasets that contain both sensing and localization data. The experiments cover both regression and classification tasks. The datasets include the open-source Wi-mans dataset, the simulated Wisimulation dataset, and our proprietary respiration sensing and localization dataset Wi-respiration. In Table \ref{table:performance_comparison}, it is shown that synergistic trained models have better performance in all three datasets.

\begin{table}[htb]
\caption{Performance Comparison on Different Datasets and Methods}
\footnotesize Note: $\uparrow$ indicates higher is better; $\downarrow$ indicates lower is better.
\label{table:performance_comparison}
\centering
\resizebox{\linewidth}{!}{%
\begin{tabular}{lccc}
\toprule
\textbf{Methods\textbackslash Metrics} & \textbf{Wi-mans} & \textbf{Simulation} & \textbf{Wi-respiration} \\
& \textbf{Acc (Classification)} & \textbf{MSE (Regression)} & \textbf{MSE (Regression)} \\
\midrule
Localization Only & 0.701 & 0.34 & 0.24 \ \  \\
Sensing Only & 0.64 & 0.13 & 0.072 \  \\
SPG-MIBO Loc &\ \  0.736 $\uparrow$ & \ \ 0.27 $\downarrow$ & \ 0.21 $\downarrow$ \\
SPG-MIBO Sen &\ \  0.682 $\uparrow$& \ \  0.09 $\downarrow$ & \ 0.056 $\downarrow$\\
\bottomrule
\end{tabular}
} 
\end{table}

\section{Discussion}
We compare the convergence behaviors of standard SGD and the SPGD with our proximal gradient operation on a relaxed mixed-integer optimization problem with binary $\{0,1\}$ constraints. The results indicate that the SPGD achieves a convergence rate comparable to that of SGD, while demonstrating improved training stability, especially when compared to SGD augmented with a penalty-based constraint formulation. In addition, the similar convergence trends of the loss curves under both SPGD and SGD-Penalty empirically confirm that SPGD achieves a convergence rate of $\mathcal{O}(1/\sqrt{T})$, which aligns well with our theoretical analysis.

Beyond convergence behavior, we evaluate the integer feasibility of the solutions. Under the same optimization framework, the binary variables obtained by SPG-MIBO exhibit a substantially smaller Integer Feasibility Gap compared to those optimized via penalty-based MIBO. This result suggests that our relaxation-tightening strategy provides a more accurate approximation to the discrete feasible space, yielding tighter solutions than traditional penalty-based approaches.

Finally, we assess the impact of the proposed optimization framework on model performance. Experiments conducted on three distinct datasets, employing identical model architectures, consistently demonstrate performance improvements in both localization and sensing tasks when trained jointly under the SPG-MIBO framework. These results primarily indicate that the modeling approach contributes to performance gains with broad applicability. Moreover, the proposed algorithm reliably ensures effective optimization across different parameter scales, further validating its robustness and generalizability.

\section{Conclusion}

In this work, we proposed a novel framework for joint localization and sensing in wireless systems by leveraging a fusion modeling approach with shared subcarrier selection. By exploiting the hierarchical structure of subcarrier information inherent to both tasks, we formulated the training problem as a MIBO problem, which naturally captures the discrete-continuous characteristics of the system. To efficiently solve this challenging problem, we developed the SPG-MIBO algorithm, a stochastic proximal gradient method under a relaxation and tightening scheme. This approach relaxes real-world soft constraints to enhance optimization efficiency while employing proximal and cutting-plane techniques to tighten the feasible region. We rigorously established the convergence rate of the algorithm, providing theoretical guarantees for its practical use. Extensive experiments on both simulated and real-world datasets demonstrated consistent performance improvements in localization and sensing tasks through our fusion modeling approach. The results validate not only the effectiveness of the proposed algorithm and model formulation but also its strong generalizability and scalability across diverse scenarios. Overall, this work presents a principled and efficient solution to the joint wireless localization and sensing problem, opening new directions for future research on complex hierarchical optimization in wireless systems.


%

\vfill
\newpage
\appendices
\section{Simulation Experiment}
\label{appendix:simulation_experiment}
Our simulation setup is based on the MIMO-OFDM channel model, which has been widely adopted in prior works focusing on model-based methods~\cite{gu2019wifi, wang2020csi}. The Channel State Information (CSI) at time \( t \) is modeled as the superposition of static-path CSI, dynamic-path CSI, and additive noise:
\begin{equation}
    H(t) = \sum_{i\in \Omega_s(t)} H_i(t) + \sum_{k\in \Omega_d(t)} H_k(t) + n(t),
\end{equation}
where \( \Omega_s(t) \) and \( \Omega_d(t) \) denote the sets of static and dynamic paths, respectively, and \( n(t) \sim \mathcal{N}(0, \sigma_f) \) represents additive Gaussian white noise.

For static paths, the signal gain \( a_i \), phase shift \( \theta_i \), and delay \( \tau_i \) are time-invariant. In contrast, for dynamic paths, the gain \( b_k(t) \), phase \( \theta_k(t) \), and delay \( \tau_k(t) \) vary over time. The CSI can thus be expressed in the following form:
\begin{equation}
\begin{aligned}
    H(t) = &\sum_{i \in \Omega_s(t)} a_i \cdot e^{j(2\pi f_i \tau_i + \theta_i)} \\
          &+ \sum_{k \in \Omega_d(t)} b_k(t) \cdot e^{j(2\pi f_k \tau_k(t) + \theta_k(t))} + n(t),
\end{aligned}
\end{equation}
where \( j \) is the imaginary unit, and \( f_i \), \( f_k \) denote the carrier frequencies of the \( i \)-th static and \( k \)-th dynamic paths, respectively.

\begin{figure}[htb]
    \centering
    \includegraphics[width=0.9\columnwidth,trim=50 70 50 50, clip]{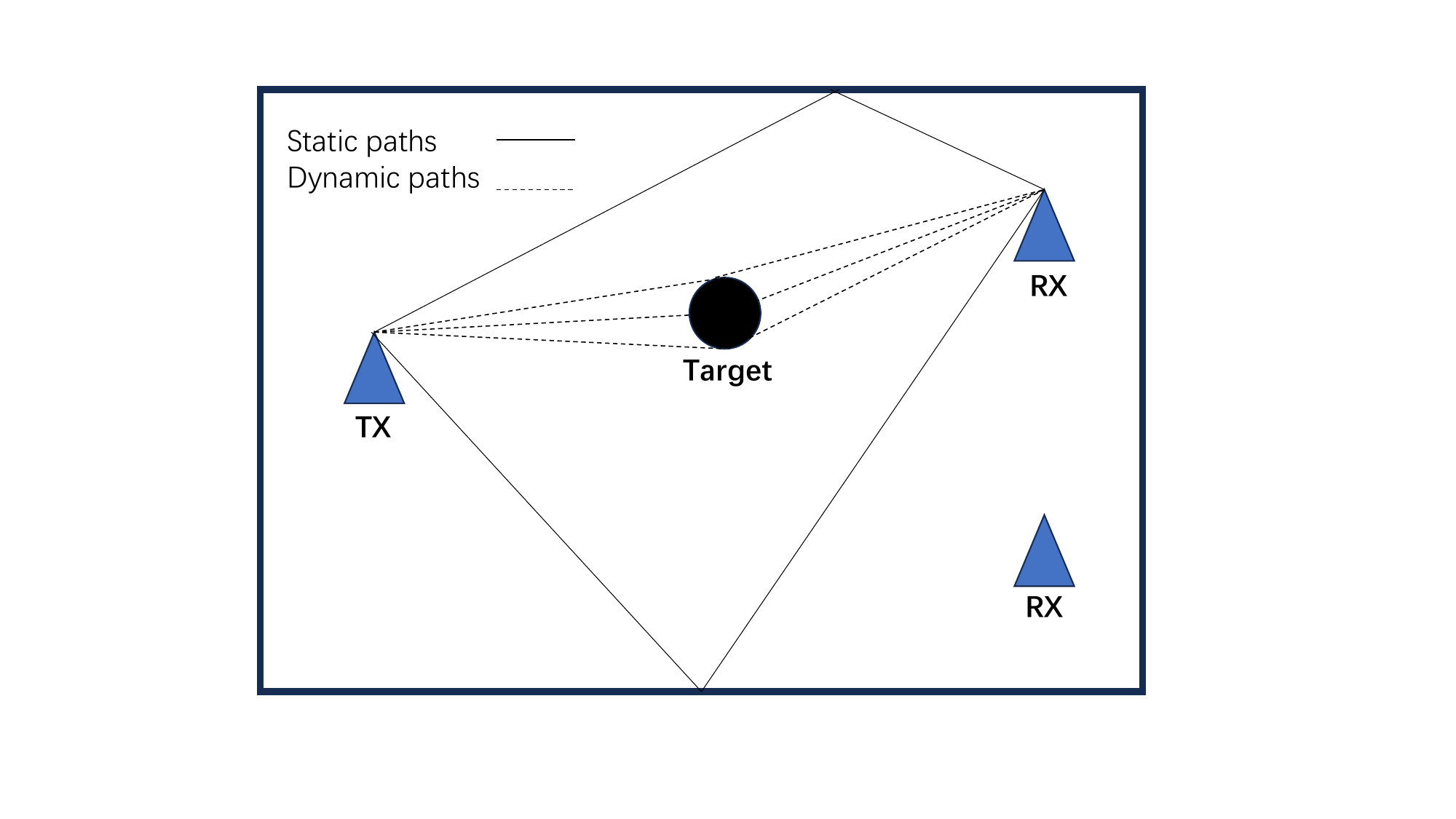} 
    \caption{Simulation Scenario Diagram}
    \label{fig:example}
\end{figure}

To simulate the physical environment, we consider a \( 4\,\mathrm{m} \times 3\,\mathrm{m} \) rectangular area with one transmitter (TX) and two receivers (RXs) (in Fig. \ref{fig:example}). For each TX/RX pair, paths reflecting only off static environmental structures (e.g., walls) are treated as static paths, while those involving interaction with the target are considered dynamic.

For static paths, the gain \( a_i \) depends on the corresponding path length \( d_i \), computed from known TX/RX positions. We assume a reference gain \( a_0 \), and apply the free-space path loss model: \( a_i = a_0 / d_i^2 \).

For dynamic paths, the path length \( d_k(t) \) depends on both the TX/RX locations and the time-varying target position \( p(t) \). Furthermore, the target's state \( s(t) \) affects the dynamic gain \( b_k(t) \). Letting \( b_0 \) be the initial dynamic gain, we define:
\begin{equation}
    b_k(t) = \frac{b_0 \cdot \gamma(s(t))}{d_k^2(t)},
\end{equation}
where \( \gamma(s(t)) \) models the target's impact on the received signal under different states. In Table \ref{tab:simulation_parameters}, we provide the relevant parameters for the simulation scenario and model configuration.
  
\begin{table}[htb]
    \centering
    \caption{Simulation Parameter Ranges}
    \renewcommand{\arraystretch}{1.2}
    \resizebox{\linewidth}{!}{
    \begin{tabular}{ccc}
        \toprule
        \textbf{Parameter} & \textbf{Range} & \textbf{Remarks} \\
        \midrule
        Number of dynamic paths \( |\Omega_d(t)| \) & \( 3 \) & More in complex environments \\
        Number of static paths \( |\Omega_s| \) & \( 3 \) & More in complex environments \\
        Gain \( a_i, b_j(t) \) & \( [-30, 0] \) dB & Higher for LoS paths, lower for dynamic paths \\
        Noise variance \( \sigma_f^2 \) & \( [-100, -70] \) dBm & Computed based on bandwidth \\
        Delay \( \tau_i, \tau_j(t) \) & \( [10, 10^4] \) ns & Computed as \( \tau = d/c \) \\
        Subcarriers number & 32-512  &  Can be dynamically adjusted\\
        Position sampling interval & 0.2 m & Uniform grid sampling\\
        \bottomrule
    \end{tabular}
    }
    \label{tab:simulation_parameters}
\end{table}

\newpage

\section{Convergence Analysis}
Assume $\mathcal{L}(\bm{u},\bm{v},\bm{w})$ is $L$-smooth where $\bm{u}\in \mathbb{R}^{n_1}$, $\bm{v}\in \mathbb{R}^{n_2}$,$\bm{w}\in [0,1]^d $ represents continuous variables, dual variables, and relaxed variables, respectively. 
The objective function with regularization term $\mathcal{G}(\bm{w})$ is:
\begin{equation}
\mathcal{F}(\bm{u},\bm{v},\bm{w}) = \mathcal{L}(\bm{u},\bm{v},\bm{w}) + \mathcal{G}(\bm{w})
\end{equation}
where $\mathcal{G}(\bm{w})$ is defined as follows:
\begin{equation}
\label{regularization function}
\begin{aligned}
     &\mathcal{G}(\bm{w}) = \|\bm{w}-\hat{\bm{x}}\|^2\\
     &s.t.\quad  \hat{\bm{x}}=\arg\min_{\bm{x}} \|\bm{w}-\bm{x}\|^2, \\
     &\qquad \  \bm{x}\in\{0,1\}^d,\bm{w}\in [0,1]^d.
\end{aligned}
\end{equation}
We define
\begin{equation}
    prox_{\eta,\mathcal{G}}(\bm{w}) = \arg\min_{\bm{x}} \left(\frac{1}{2\eta} \|\bm{x} - \bm{w}\|^2 + \mathcal{G}(\bm{x}) \right).
\end{equation}

The above proximal operation is equivalent to the following closed-form expression:
\begin{equation}
    prox_{\eta,\mathcal{G}}(\bm{w}) = [prox_{\eta,\mathcal{G}}(w_1), prox_{\eta,\mathcal{G}}(w_2), \ldots, prox_{\eta,\mathcal{G}}(w_d)],
\end{equation}
\begin{equation}
    prox_{\eta,\mathcal{G}}(w_i) = \begin{cases}
        \frac{w_i/\eta}{\frac{1}{\eta} + 2} & , w_i \in [0, \frac{1}{2}] \\
        \frac{w_i/\eta + 2}{\frac{1}{\eta} + 2} & , w_i\in (\frac{1}{2}, 1].
    \end{cases}
\end{equation}
When variables are updated as:
\begin{equation}
\begin{aligned}
    \bm{u}_{k+1} &= \bm{u}_k - \eta_k \nabla_{\bm{u}} \mathcal{L}(\bm{u}_k,\bm{v}_k,\bm{w}_k; \xi_k) \\
    \bm{v}_{k+1} &= \bm{v}_k + \eta_k \nabla_{\bm{v}} \mathcal{L}(\bm{u}_k,\bm{v}_k,\bm{w}_k; \xi_k)\\
    \bar{\bm{w}}_{k+1} &= \bar{\bm{w}}_k - \eta_k \nabla_{\bm{w}} \mathcal{L}(\bar{\bm{w}}_k, \bm{v}_k, \bm{w}_k; \xi_k)\\
    \bm{w}_{k+1}&=prox_{\eta_k,\mathcal{G}}(\bar{\bm{w}}_{k+1}),
\end{aligned}
\end{equation}
the L-smooth property follows:
\begin{equation}
\resizebox{\columnwidth}{!}{$
    \begin{aligned}
        & \mathcal{L}(\bm{u}_{k+1}, \bm{v}_{k+1}, \bm{w}_{k+1}) 
        \\   \leq  & \mathcal{L}(\bm{u}_k, \bm{v}_k, \bm{w}_k) 
         + 
        \begin{pmatrix}
            \nabla_{\bm{u}} \mathcal{L}(\bm{u}_k, \bm{v}_k, \bm{w}_k) \\
            \nabla_{\bm{v}} \mathcal{L}(\bm{u}_k, \bm{v}_k, \bm{w}_k) \\
            \nabla_{\bm{w}} \mathcal{L}(\bm{u}_k, \bm{v}_k, \bm{w}_k)
        \end{pmatrix}^T 
        \begin{pmatrix}
            \bm{u}_{k+1} - \bm{u}_k \\
            \bm{v}_{k+1} - \bm{v}_k \\
            \bm{w}_{k+1} - \bm{w}_k
        \end{pmatrix} \\
         &+ \frac{L}{2} \left\| \begin{pmatrix}
            \bm{u}_{k+1} - \bm{u}_k \\
            \bm{v}_{k+1} - \bm{v}_k \\
            \bm{w}_{k+1} - \bm{w}_k
        \end{pmatrix} \right\|^2.
    \end{aligned}
$}
\end{equation}

\subsection{Proof of Theorem 1 (Convergence of $\bm{w}$)}
Since the updating of $\bm{w}$ satisfies: 
\begin{equation}
\resizebox{\columnwidth}{!}{$
\begin{aligned}
&\bm{w}_{k+1} \\
=& \arg \min_{\bm{w}} \left(\mathcal{G}(\bm{w}) + \frac{1}{2\eta_k}\|\bm{w} - (\bm{w}_k - \eta_k \nabla_{\bm{w}} \mathcal{L}(\bm{u}_k, \bm{v}_k, \bm{w}_k; \bm{\xi}_k))\|^2\right),
\end{aligned}
$}
\end{equation}
where $\nabla_{\bm{w}} \mathcal{L}(\bm{u}_k, \bm{v}_k, \bm{w}_k; \bm{\xi}_k) $ is the stochastic gradient with samples $\bm{\xi}_k$.

With the L-smooth assumption, we have
\begin{equation}
\resizebox{\columnwidth}{!}{$
\label{L-smooth equality}
\begin{aligned}
    &\mathcal{L}(\bm{u}_{k+1}, \bm{v}_{k+1}, \bm{w}_{k+1}) \\
    \leq & \mathcal{L}(\bm{u}_k, \bm{v}_k, \bm{w}_k) 
     + 
    \begin{pmatrix}
        \nabla_{\bm{u}} \mathcal{L}(\bm{u}_k, \bm{v}_k, \bm{w}_k) \\
        \nabla_{\bm{v}} \mathcal{L}(\bm{u}_k, \bm{v}_k, \bm{w}_k) \\
        \nabla_{\bm{w}} \mathcal{L}(\bm{u}_k, \bm{v}_k, \bm{w}_k)
    \end{pmatrix}^T 
    \begin{pmatrix}
        \bm{u}_{k+1} - \bm{u}_k \\
        \bm{v}_{k+1} - \bm{v}_k \\
        \bm{w}_{k+1} - \bm{w}_k
    \end{pmatrix} \\
     &+ \frac{L}{2} \left\| \begin{pmatrix}
        \bm{u}_{k+1} - \bm{u}_k \\
        \bm{v}_{k+1} - \bm{v}_k \\
        \bm{w}_{k+1} - \bm{w}_k
    \end{pmatrix} \right\|^2.
\end{aligned}
$}
\end{equation}
Since $\bm{w}_{k+1}$ is the minimizer, we have
\begin{equation}
\label{mimimizer equality}
\resizebox{\columnwidth}{!}{$
\begin{aligned}
    &\mathcal{G}(\bm{w}_{k+1}) + \frac{1}{2\eta_k} \|\bm{w}_{k+1} - (\bm{w}_k - \eta_k \nabla_{\bm{w}_t} \mathcal{L}(\bm{u}_k, \bm{v}_k, \bm{w}_k; \boldsymbol{\xi}_t))\|^2 \\
    \leq &\mathcal{G}(\bm{w}_k) + \frac{1}{2\eta_k} \|\bm{w}_k - (\bm{w}_k - \eta_k \nabla_{\bm{w}_t} \mathcal{L}(\bm{u}_k, \bm{v}_k, \bm{w}_k; \boldsymbol{\xi}_t))\|^2.
\end{aligned}
$}
\end{equation}
Combining the above (\ref{L-smooth equality}) and (\ref{mimimizer equality}) leads to the following inequality:
\begin{equation}
\resizebox{\columnwidth}{!}{$
\begin{aligned}
& \mathcal{L}(\bm{u}_{k+1}, \bm{v}_{k+1}, \bm{w}_{k+1}) + \mathcal{G}(\bm{w}_{k+1})\\
&+ \frac{1}{2\eta_k} \|\bm{w}_{k+1} - (\bm{w}_k - \eta_k \nabla \mathcal{L}(\bm{u}_k, \bm{v}_k, \bm{w}_k; \bm{\xi}_t))\|^2 \\
\leq & \mathcal{L}(\bm{u}_k, \bm{v}_k, \bm{w}_k) + \mathcal{G}(\bm{w}_k) \\
&+ \nabla_{\bm{w}} \mathcal{L}(\bm{u}_k, \bm{v}_k, \bm{w}_k)^T (\bm{w}_{k+1} - \bm{w}_k) + \frac{L}{2} \|\bm{w}_{k+1} - \bm{w}_k\|^2 \\
 & +\frac{\eta_k}{2} \|\nabla \mathcal{L}(\bm{u}_k, \bm{v}_k, \bm{w}_k; \bm{\xi}_t)\|^2+\mathcal{R}_k(\bm{u},\bm{v}),
\end{aligned}
$}
\end{equation}
where $\mathcal{R}_k(\bm{u},\bm{v})$ is given as:
\begin{equation}
\begin{aligned}
    &\mathcal{R}_k(\bm{u},\bm{v})\\
    =&
    \begin{bmatrix}
    \nabla_{\bm{u}} \mathcal{L}(\bm{u}_k, \bm{v}_k, \bm{u}_k) \\
    \nabla_{\bm{v}} \mathcal{L}(\bm{u}_k, \bm{v}_k, \bm{v}_k)
    \end{bmatrix}^T
    \begin{bmatrix}
    \bm{u}_{k+1} - \bm{u}_k \\
    \bm{v}_{k+1} - \bm{v}_k
    \end{bmatrix}\\
&+ \frac{L}{2} \left( \|\bm{u}_{k+1} - \bm{u}_k\|^2 + \|\bm{v}_{k+1} - \bm{v}_k\|^2 \right)\\
= & \left(- \eta_k + \frac{L}{2} \eta_k^2\right) \|\nabla_{\bm{u}} \mathcal{L}(\bm{u}_k, \bm{v}_k, \bm{w}_k; \xi_k)\|^2 \\
&+ \left(\eta_k + \frac{L}{2} \eta_k^2\right) \|\nabla_{\bm{v}} \mathcal{L}(\bm{u}_k, \bm{v}_k, \bm{w}_k; \xi_k)\|^2.
\end{aligned}
\end{equation}
Replacing $\mathcal{L}(\bm{u},\bm{v},\bm{w})+\mathcal{G}(\bm{w})$ with $\mathcal{F}(\bm{u},\bm{v},\bm{w})$ and expanding the left side of the inequality, we have
\begin{equation}
    \begin{aligned}
    & \mathcal{F}(\bm{u}_{k+1}, \bm{v}_{k+1}, \bm{w}_{k+1}) \\
    &+ \frac{1}{2\eta_k} \|\bm{w}_{k+1} - (\bm{w}_k - \eta_k \nabla_{\bm{w}} \mathcal{L}(\bm{u}_k, \bm{v}_k, \bm{w}_k; \bm{\xi}_k))\|^2 \\
    \leq & \mathcal{F}(\bm{u}_k, \bm{v}_k, \bm{w}_k) + \nabla_{\bm{w}} \mathcal{L}(\bm{u}_k, \bm{v}_k, \bm{w}_k)^T (\bm{w}_{k+1} - \bm{w}_k)\\
    &+ \frac{L}{2} \|\bm{w}_{k+1} - \bm{w}_k\|^2 \\
    & + \frac{\eta_k}{2} \|\nabla_{\bm{w}} \mathcal{L}(\bm{u}_k, \bm{v}_k, \bm{w}_k; \bm{\xi}_t)\|^2 + \mathcal{R}_k(\bm{u}, \bm{v}).
    \end{aligned}
\end{equation}
Thus,
\begin{equation}
    \begin{aligned}
    &\mathcal{F}(\bm{u}_{k+1}, \bm{v}_{k+1}, \bm{w}_{k+1})+\frac{1}{2\eta_k}\|\bm{w}_{k+1} - \bm{w}_k\|^2 \\
    &+ \nabla_{\bm{w}} \mathcal{L}(\bm{u}_k, \bm{v}_k, \bm{w}_k; \bm{\xi}_k)^T (\bm{w}_{k+1} - \bm{w}_k)\\
    \leq & \mathcal{F}(\bm{u}_k, \bm{v}_k, \bm{w}_k)  + \frac{L}{2} \|\bm{w}_{k+1} - \bm{w}_k\|^2\\
    &+ \nabla_{\bm{w}} \mathcal{L}(\bm{u}_k, \bm{v}_k, \bm{w}_k)^T (\bm{w}_{k+1} - \bm{w}_k)  + \mathcal{R}_k(\bm{u}, \bm{v}).
    \end{aligned}
\end{equation}
Taking the expectation of the stochastic gradient as:
\begin{equation}
\begin{aligned}
        & \mathcal{F}(\bm{u}_{k+1}, \bm{v}_{k+1},\bm{w}_{k+1})-\mathcal{F}(\bm{u}_k, \bm{v}_k, \bm{w}_k)\\
        &+\mathbb{E}[\nabla_{\bm{w}} \mathcal{L}(\bm{u}_k, \bm{v}_k, \bm{w}_k; \bm{\xi}_k)^T ](\bm{w}_{k+1} - \bm{w}_k)\\
        \leq &\left(\frac{L}{2} - \frac{1}{2\eta_k}\right) \|\bm{w}_{k+1} - \bm{w}_k\|^2\\
        &+\nabla_{\bm{w}} \mathcal{L}(\bm{u}_k, \bm{v}_k, \bm{w}_k)^T (\bm{w}_{k+1} - \bm{w}_k) \\
        &+\mathbb{E}[\mathcal{R}_k(\bm{u}, \bm{v})].
\end{aligned}
\end{equation}
Since
\begin{equation}
    \begin{aligned}
        &\mathbb{E}[\nabla_{\bm{w}} \mathcal{L}(\bm{u}_k, \bm{v}_k, \bm{w}_k; \bm{\xi}_k)]=\nabla_{\bm{w}} \mathcal{L}(\bm{u}_k, \bm{v}_k, \bm{w}_k).
    \end{aligned}
\end{equation}
We have the following:
\begin{equation}
\begin{aligned}
     &\mathcal{F}(\bm{u}_{k+1}, \bm{v}_{k+1},\bm{w}_{k+1})-\mathcal{F}(\bm{u}_k, \bm{v}_k, \bm{w}_k) \\
    \leq &\left(\frac{L}{2} - \frac{1}{2\eta_k}\right) \|\bm{w}_{k+1} - \bm{w}_k\|^2 \\
    &+\mathbb{E}[\mathcal{R}_k(\bm{u}, \bm{v})].
\end{aligned}
\end{equation}
By summing up $T$ steps, we have
\begin{equation}
\label{T steps equality of w}
\begin{aligned} 
&\sum_{k=1}^T \left(\frac{1}{2\eta_k}-\frac{L}{2} \right)\mathbb{E}[\|\bm{w}_{k+1} - \bm{w}_k\|^2] \\
    \leq & \mathcal{F}(\bm{u}_{1}, \bm{v}_{1}, \bm{w}_{1})-\mathcal{F}(\bm{u}_{T+1}, \bm{v}_{T+1}, \bm{w}_{T+1})\\
    &+\sum_{k=1}^T \mathbb{E}[\mathcal{R}_k(\bm{u}, \bm{v})].
\end{aligned}
\end{equation}
Taking step-size $\eta_k = \frac{\eta}{\sqrt{k}}$ and $\frac{1}{2\eta_k}-\frac{L}{2}>0$, we have
\begin{equation}
\label{sums with stepsize}
\begin{aligned}
   \sum_{k=1}^T \left( \frac{\sqrt{k}}{2\eta}-\frac{L}{2} \right) &\approx \frac{T^{3/2}+1}{3\eta} -\frac{LT}{2}   \\
   \sum_{k=1}^T\frac{1}{k} &\approx \log(T)\\
   \sum_{k=1}^T\frac{1}{\sqrt{k}} &\approx 2\sqrt{T}.
\end{aligned}
\end{equation}
Denoting $\sigma_{\bm{u}},\sigma_{\bm{v}},\sigma_{\bm{w}}$ as the variance of the stochastic gradient of $\bm{u},\bm{v},\bm{w}$, we have
\begin{equation}
\begin{aligned}
    &\mathbb{E}[\|\nabla_{\bm{u}} \mathcal{L}(\bm{u}_k, \bm{v}_k, \bm{w}_k; \xi_k)\|^2] \\
    \leq & \mathbb{E}[\|\nabla_{\bm{u}} \mathcal{L}(\bm{u}_k, \bm{v}_k, \bm{w}_k; \xi_k)-\nabla_{\bm{u}} \mathcal{L}(\bm{u}_k, \bm{v}_k, \bm{w}_k)\|^2\\
    &+\|\nabla_{\bm{u}} \mathcal{L}(\bm{u}_k, \bm{v}_k, \bm{w}_k)\|^2]\\
\leq & \sigma_{\bm{u}}^2 + \mathbb{E}[\|\nabla_{\bm{u}} \mathcal{L}(\bm{u}_k, \bm{v}_k, \bm{w}_k)\|^2].
\end{aligned}
\end{equation}
Thus, we have
\begin{equation}
\label{expectation of R(u,v)}
\begin{aligned}
&\sum_{k=1}^T\mathbb{E}[\mathcal{R}_k(\bm{u}, \bm{v})] \\
\approx &  \left(\frac{L\eta^2}{2} \log(T)-2\eta\sqrt{T}\right)\left(\mathbb{E}[\|\nabla_{\bm{u}} \mathcal{L}(\bm{u}, \bm{v}, \bm{w})\|^2]+\sigma_{\bm{u}}\right) \\
& +\left(\frac{L\eta^2}{2} \log(T)+2\eta\sqrt{T}\right) \left(\mathbb{E}[\| \nabla_{\bm{v}} \mathcal{L}(\bm{u}, \bm{v}, \bm{w})\|^2 ]+\sigma_{\bm{v}}\right).
\end{aligned}
\end{equation}
Combining (\ref{T steps equality of w}), (\ref{sums with stepsize}) and (\ref{expectation of R(u,v)}), we have
\begin{equation}
\resizebox{\columnwidth}{!}{$
\begin{aligned}
    &\frac{1}{T} \sum_{k=1}^T \mathbb{E}[\|\bm{w}_{k+1} - \bm{w}_k\|^2]\\
    \leq & \left( \frac{6\eta}{2\sqrt{T}+\frac{2}{T}-3\eta L}\right)\left(\mathcal{F}(\bm{u}_{1}, \bm{v}_{1}, \bm{w}_{1})-\mathcal{F}(\bm{u}_{T+1}, \bm{v}_{T+1}, \bm{w}_{T+1})\right)\\
   &+\left(\frac{L\eta^2\log(T)}{2T} -\frac{2\eta}{\sqrt{T}}\right)\left(\mathbb{E}[\|\nabla_{\bm{u}} \mathcal{L}(\bm{u}, \bm{v}, \bm{w})\|^2]+\sigma_{\bm{u}}\right)\\
   &+\left(\frac{L\eta^2\log(T)}{2T} + \frac{2\eta}{\sqrt{T}}\right)\left(\mathbb{E}[\|\nabla_{\bm{v}} \mathcal{L}(\bm{u}, \bm{v}, \bm{w})\|^2]+\sigma_{\bm{v}}\right).
\end{aligned}
$}
\end{equation}
With the assumptions that the gradient is bounded, such that $\|\nabla_{\bm{u}} \mathcal{L}(\bm{u}_k, \bm{v}_k, \bm{w}_k)\|\leq G_{\bm{u}}$ and $\|\nabla_{\bm{v}} \mathcal{L}(\bm{u}_k, \bm{v}_k, \bm{w}_k)\|\leq G_{\bm{v}}$,$\mathcal{F}$ has a lower bound, such that $
\mathcal{F}^*\leq \mathcal{F}(\bm{u}_{T+1}, \bm{v}_{T+1}, \bm{w}_{T+1})$, we have
\begin{equation}
\begin{aligned}
    &\frac{1}{T} \sum_{k=1}^T \mathbb{E}[\|\bm{w}_{k+1} - \bm{w}_k\|^2]\\
    \leq & \left( \frac{6\eta}{2\sqrt{T}+\frac{2}{T}-3\eta L}\right)\left(\mathcal{F}(\bm{u}_{1}, \bm{v}_{1}, \bm{w}_{1})-\mathcal{F}^*\right)\\
   &+\left(\frac{L\eta^2\log(T)}{2T} -\frac{2\eta}{\sqrt{T}}\right)\left(G_{\bm{u}}^2+\sigma_{\bm{u}}\right)\\
   &+\left(\frac{L\eta^2\log(T)}{2T} + \frac{2\eta}{\sqrt{T}}\right)\left(G_{\bm{v}}^2+\sigma_{\bm{v}}\right).
\end{aligned}
\end{equation}
Thus, $\bm{w}$ has a convergence rate of $\mathcal{O}(\frac{1}{\sqrt{T}})$.
\subsection{Proof of Theorem 2 (Convergence of $\bm{u}$)}
Based on the L-smooth assumption we have
\begin{equation}
\begin{aligned}
    &\mathcal{L}(\bm{u}_{k+1},\bm{v}_{k+1},\bm{w}_{k+1}) \\
    \leq & \mathcal{L}(\bm{u}_k,\bm{v}_k,\bm{w}_k) + \nabla_{\bm{u}} \mathcal{L}(\bm{u}_k,\bm{v}_k,\bm{w}_k)^\top (\bm{u}_{k+1}-\bm{u}_k) \\
    &+ \frac{L}{2} \|\bm{u}_{k+1}-\bm{u}_k\|^2+\mathcal{R}_k(\bm{v},\bm{w}),
\end{aligned}
\end{equation}
where $\mathcal{R}_k(\bm{v},\bm{w})$ is as follows:
\begin{equation}
\begin{aligned}
    \mathcal{R}_k(\bm{v},\bm{w})=&
    \begin{bmatrix}
    \nabla_{\bm{v}} \mathcal{L}(\bm{v}_k, \bm{w}_k, \bm{v}_k) \\
    \nabla_{\bm{w}} \mathcal{L}(\bm{v}_k, \bm{w}_k, \bm{w}_k)
    \end{bmatrix}^T
    \begin{bmatrix}
    \bm{v}_{k+1} - \bm{v}_k \\
    \bm{w}_{k+1} - \bm{w}_k
    \end{bmatrix}\\
&+ \frac{L}{2} \left( \|\bm{v}_{k+1} - \bm{v}_k\|^2 + \|\bm{w}_{k+1} - \bm{w}_k\|^2 \right)\\
= & \left(\eta_k + \frac{L}{2} \eta_k^2\right) \|\nabla_{\bm{v}} \mathcal{L}(\bm{v}_k, \bm{w}_k, \bm{w}_k; \xi_k)\|^2 \\
&+  \nabla_{\bm{w}} \mathcal{L}(\bm{v}_k, \bm{w}_k, \bm{w}_k)^T(\bm{w}_{k+1} - \bm{w}_k)\\
&+\frac{L}{2}\|\bm{w}_{k+1} - \bm{w}_k\|^2.
\end{aligned}
\end{equation}
With $\bm{u}_{k+1} = \bm{u}_k - \eta_k \nabla_{\bm{u}} \mathcal{L}(\bm{u}_k,\bm{v}_k,\bm{w}_k; \xi_k)$, we have
\begin{equation}
\resizebox{\columnwidth}{!}{$
\begin{aligned}
       &\mathcal{L}(\bm{u}_{k+1},\bm{v}_{k+1},\bm{w}_{k+1})\\
       \leq &\mathcal{L}(\bm{u}_k,\bm{v}_k,\bm{w}_k) - \eta_k \nabla_{\bm{u}} \mathcal{L}(\bm{u}_k,\bm{v}_k,\bm{w}_k)^\top \nabla_{\bm{u}} \mathcal{L}(\bm{u}_k,\bm{v}_k,\bm{w}_k; \xi_k)  \\
       &+ \frac{L}{2} \eta_k^2\| \nabla_{\bm{u}} \mathcal{L}(\bm{u}_k,\bm{v}_k,\bm{w}_k; \xi_k)\|^2 +\mathcal{R}_k(\bm{v},\bm{w}).
\end{aligned}
$}
\end{equation}
Taking the expectation of the stochastic gradient, we have
\begin{equation}
\label{inequality with expection of gradient u}
\begin{aligned}
     &\mathcal{L}(\bm{u}_{k+1},\bm{v}_{k+1},\bm{w}_{k+1})-\mathcal{L}(\bm{u}_k,\bm{v}_k,\bm{w}_k) \\
     \leq &\frac{L}{2} \eta_k^2\mathbb{E}[\| \nabla_{\bm{u}} \mathcal{L}(\bm{u}_k,\bm{v}_k,\bm{w}_k; \xi_k)\|^2]\\
     &-\eta_k \nabla_{\bm{u}} \mathcal{L}(\bm{u}_k,\bm{v}_k,\bm{w}_k)^\top \mathbb{E}[\nabla_{\bm{u}} \mathcal{L}(\bm{u}_k,\bm{v}_k,\bm{w}_k; \xi_k) ]\\
     &+\mathbb{E}[\mathcal{R}_k(\bm{v},\bm{w})].
\end{aligned}
\end{equation}
According to the assumption of the unbiased estimation of the stochastic gradient and defining the variance of the stochastic gradient of $\bm{u}$ as $\sigma_{\bm{u}}$, we have
\begin{equation}
\begin{aligned}
    \mathbb{E}[\nabla_{\bm{u}} \mathcal{L}(\bm{u}_k, \bm{v}_k, \bm{w}_k; \xi_k)] &= \nabla_{\bm{u}} \mathcal{L}(\bm{u}_k, \bm{v}_k, \bm{w}_k)\\
    \mathbb{E}[\nabla_{\bm{v}} \mathcal{L}(\bm{u}_k, \bm{v}_k, \bm{w}_k; \xi_k)] &= \nabla_{\bm{v}} \mathcal{L}(\bm{u}_k, \bm{v}_k, \bm{w}_k),
\end{aligned}
\end{equation}
\begin{equation}
\begin{aligned}
    &\mathbb{E}[\|\nabla_{\bm{u}} \mathcal{L}(\bm{u}_k, \bm{v}_k, \bm{w}_k; \xi_k)\|^2]\\
    =&\mathbb{E}[\|(\nabla_{\bm{u}} \mathcal{L}(\bm{u}_k, \bm{v}_k, \bm{w}_k; \xi_k)-\nabla_{\bm{u}} \mathcal{L}(\bm{u}_k, \bm{v}_k, \bm{w}_k))\\
    &+\nabla_{\bm{u}} \mathcal{L}(\bm{u}_k, \bm{v}_k, \bm{w}_k)\|^2]\\
    \leq& \sigma_{\bm{u}}^2 + \mathbb{E}[\|\nabla_{\bm{u}} \mathcal{L}(\bm{u}_k, \bm{v}_k, \bm{w}_k)\|^2].
\end{aligned}
\end{equation}
Thus, by summing up (\ref{inequality with expection of gradient u}) for $T$ steps, we have
\begin{equation}
\label{inequality summed over k}
\begin{aligned}
     &\mathcal{L}(\bm{u}_{T+1},\bm{v}_{T+1},\bm{w}_{T+1})-\mathcal{L}(\bm{u}_1,\bm{v}_1,\bm{w}_1) \\
     \leq &\sum_{k=1}^T(\frac{L}{2} \eta_k^2-\eta_k)\mathbb{E}[\|\nabla_{\bm{u}} \mathcal{L}(\bm{u}_k, \bm{v}_k, \bm{w}_k)\|^2]\\
     &+\frac{L}{2} \eta_k^2\sigma_{\bm{u}}^2+\sum_{k=1}^T \mathbb{E}[\mathcal{R}_k(\bm{v},\bm{w})].
\end{aligned}
\end{equation}
Taking stepsize $\eta_k=\frac{\eta}{\sqrt{k}}$, we have
\begin{equation}
\begin{aligned}
     &\mathcal{L}(\bm{u}_{T+1},\bm{v}_{T+1},\bm{w}_{T+1})-\mathcal{L}(\bm{u}_1,\bm{v}_1,\bm{w}_1) \\
     \leq &\sum_{k=1}^T \left(\frac{L\eta^2}{2k} - \frac{\eta}{\sqrt{k}}\right) \mathbb{E}[\|\nabla_{\bm{u}} \mathcal{L}(\bm{u}_k, \bm{v}_k, \bm{w}_k)\|^2]\\
     &+ \frac{L\eta^2}{2} \sum_{k=1}^T \frac{\sigma_{\bm{u}}^2}{k} + \sum_{k=1}^T \mathbb{E}[\mathcal{R}_k(\bm{v},\bm{w})],
\end{aligned}
\end{equation}
where
\begin{equation}
\begin{aligned}
    &\sum_{k=1}^T \mathbb{E}[\mathcal{R}_k(\bm{v},\bm{w})]\\
    =&\frac{LT}{2}\mathbb{E}[\|\bm{w}_{k+1} - \bm{w}_k\|^2]\\
     &+\sum_{k=1}^T \left(\frac{L\eta^2}{2k} + \frac{\eta}{\sqrt{k}}\right)\mathbb{E}[\|\nabla_{\bm{v}} \mathcal{L}(\bm{u}_k, \bm{v}_k, \bm{w}_k)\|^2]\\
     &+\sum_{k=1}^T\mathbb{E}[\nabla_{\bm{w}} \mathcal{L}(\bm{v}_k, \bm{w}_k, \bm{w}_k)^T(\bm{w}_{k+1} - \bm{w}_k)]
\end{aligned}
\end{equation}
With $\sum_{k=1}^T \frac{1}{k}\approx \log(T),\sum_{k=1}^T \frac{1}{\sqrt{k}}\approx 2\sqrt{T}$, we have
\begin{equation}
    \begin{aligned}
        \frac{L\eta^2}{2k} - \frac{\eta}{\sqrt{k}} &\approx \frac{L\eta^2\log(T)}{2} - 2\eta\sqrt{T}\\
       \frac{L\eta^2}{2k} + \frac{\eta}{\sqrt{k}} &\approx \frac{L\eta^2\log(T)}{2} + 2\eta\sqrt{T}.
    \end{aligned}
\end{equation}
Thus, we have :
\begin{equation}
    \begin{aligned}
        &\frac{1}{T}\sum_{k=1}^T\mathbb{E}[\|\nabla_{\bm{u}} \mathcal{L}(\bm{u}_k, \bm{v}_k, \bm{w}_k)\|^2]\\
        \leq &  \frac{\mathcal{L}(\bm{u}_1,\bm{v}_1,\bm{w}_1)-\mathcal{L}(\bm{u}_{T+1},\bm{v}_{T+1},\bm{w}_{T+1})}{\mathcal{C}_{\eta,L}(T)}+\frac{L\eta^2\sigma_{\bm{u}}^2\log(T)}{\mathcal{C}_{\eta,L}(T)}\\
        &+\frac{LT\mathbb{E}[\|\bm{w}_{k+1} - \bm{w}_k\|^2]}{2\mathcal{C}_{\eta,L}(T)}\\
        &+\frac{(2\eta\sqrt{T}+\frac{L\eta^2\log(T)}{2})\mathbb{E}[\|\nabla_{\bm{v}} \mathcal{L}(\bm{u}_k, \bm{v}_k, \bm{w}_k)\|^2]}{\mathcal{C}_{\eta,L}(T)}\\
        &+\frac{\sum_{k=1}^T\mathbb{E}[\nabla_{\bm{w}} \mathcal{L}(\bm{v}_k, \bm{w}_k, \bm{w}_k)^T(\bm{w}_{k+1} - \bm{w}_k)]}{\mathcal{C}_{\eta,L}(T)},
    \end{aligned}
\end{equation}
where $\mathcal{C}_{\eta,L}(T) = T(2\eta\sqrt{T}-\frac{L\eta^2\log(T)}{2})$.

With the assumption that the gradient is bounded, such that $\|\nabla_{\bm{v}} \mathcal{L}(\bm{u}_k, \bm{v}_k, \bm{w}_k)\|\leq G_{\bm{v}}$ and $\|\nabla_{\bm{w}} \mathcal{L}(\bm{u}_k, \bm{v}_k, \bm{w}_k)\|\leq G_{\bm{w}}$, we have the following:
\begin{equation}
\begin{aligned}
        &\frac{(2\eta\sqrt{T}+\frac{L\eta^2\log(T)}{2})\mathbb{E}[\|\nabla_{\bm{v}} \mathcal{L}(\bm{u}_k, \bm{v}_k, \bm{w}_k)\|^2]}{\mathcal{C}_{\eta,L}(T)}\\
        \leq   &\frac{(2\eta\sqrt{T}+\frac{L\eta^2\log(T)}{2})G_{\bm{v}}^2}{\mathcal{C}_{\eta,L}(T)},\\
        &\frac{\sum_{k=1}^T\nabla_{\bm{w}} \mathcal{L}(\bm{v}_k, \bm{w}_k, \bm{w}_k)^T(\bm{w}_{k+1} - \bm{w}_k)}{\mathcal{C}_{\eta,L}(T)}\\
        \leq &\frac{G_{\bm{w}}\sum_{k=1}^T \mathbb{E}[\|\bm{w}_{k+1} - \bm{w}_k\|]}{\mathcal{C}_{\eta,L}(T)}.
\end{aligned}
\end{equation}
Since $\bm{w}_k$ is convergent, then $ \mathbb{E}[\|\bm{w}_{k+1} - \bm{w}_k\|]$ is bounded, assuming $\mathbb{E}[\|\bm{w}_{k+1} - \bm{w}_k\|\leq M_{\bm{w}}$ thus, we have the following:
\begin{equation}
    \begin{aligned}
        &\frac{1}{T}\sum_{k=1}^T\mathbb{E}[\|\nabla_{\bm{u}} \mathcal{L}(\bm{u}_k, \bm{v}_k, \bm{w}_k)\|^2]\\
        \leq &  \frac{\mathcal{L}(\bm{u}_1,\bm{v}_1,\bm{w}_1)-\mathcal{L}(\bm{u}_{T+1},\bm{v}_{T+1},\bm{w}_{T+1})}{\mathcal{C}_{\eta,L}(T)}+\frac{L\eta^2\sigma_{\bm{u}}^2\log(T)}{\mathcal{C}_{\eta,L}(T)}\\
        &+\frac{LTM_{\bm{w}}^2}{2\mathcal{C}_{\eta,L}(T)}+\frac{(2\eta\sqrt{T}+\frac{L\eta^2\log(T)}{2})G_{\bm{v}}}{\mathcal{C}_{\eta,L}(T)}+\frac{G_{\bm{w}}TM_{\bm{w}}}{\mathcal{C}_{\eta,L}(T)}.
    \end{aligned}
\end{equation}
Since $\mathcal{C}_{\eta,L}(T) = T(2\eta\sqrt{T}-\frac{L\eta^2\log(T)}{2})=\mathcal{O}(T^{1.5})$, we have
\begin{equation}
\begin{aligned}
    &\max\{\frac{1}{\mathcal{C}_{\eta,L}(T)},\frac{L\eta^2\sigma_{\bm{u}}^2\log(T)}{\mathcal{C}_{\eta,L}(T)},\frac{LTM_{\bm{w}}^2}{2\mathcal{C}_{\eta,L}(T)},\\
    &\frac{(2\eta\sqrt{T}+\frac{L\eta^2\log(T)}{2})G_{\bm{v}}}{\mathcal{C}_{\eta,L}(T)},\frac{G_{\bm{w}}TM_{\bm{w}}}{\mathcal{C}_{\eta,L}(T)}\}\\
    = & \mathcal{O}(\frac{1}{\sqrt{T}}).
\end{aligned}
\end{equation}
Thus, $\bm{u}$ as a convergence rate of $\mathcal{O}(\frac{1}{\sqrt{T}})$.
\subsection{Proof of Theorem 3 (Convergence of $\bm{v}$)}
Based on the L-smooth assumption we have
\begin{equation}
\begin{aligned}
    &\mathcal{L}(\bm{u}_{k+1},\bm{v}_{k+1},\bm{w}_{k+1}) \\
    \leq & \mathcal{L}(\bm{u}_k,\bm{v}_k,\bm{w}_k) + \nabla_{\bm{v}} \mathcal{L}(\bm{u}_k,\bm{v}_k,\bm{w}_k)^\top (\bm{v}_{k+1}-\bm{v}_k) \\
    &+ \frac{L}{2} \|\bm{v}_{k+1}-\bm{v}_k\|^2 + \mathcal{R}_k(\bm{u}, \bm{w}),
\end{aligned}
\end{equation}
where $\mathcal{R}_k(\bm{u},\bm{w})$ is as follows:
\begin{equation}
\begin{aligned}
    \mathcal{R}_k(\bm{u},\bm{w})=&
    \begin{bmatrix}
    \nabla_{\bm{u}} \mathcal{L}(\bm{u}_k, \bm{w}_k, \bm{u}_k) \\
    \nabla_{\bm{w}} \mathcal{L}(\bm{u}_k, \bm{w}_k, \bm{w}_k)
    \end{bmatrix}^T
    \begin{bmatrix}
    \bm{u}_{k+1} - \bm{u}_k \\
    \bm{w}_{k+1} - \bm{w}_k
    \end{bmatrix}\\
&+ \frac{L}{2} \left( \|\bm{u}_{k+1} - \bm{u}_k\|^2 + \|\bm{w}_{k+1} - \bm{w}_k\|^2 \right)\\
= & \left(-\eta_k + \frac{L}{2} \eta_k^2\right) \|\nabla_{\bm{u}} \mathcal{L}(\bm{u}_k, \bm{w}_k, \bm{w}_k; \xi_k)\|^2 \\
& +  \nabla_{\bm{w}} \mathcal{L}(\bm{u}_k, \bm{w}_k, \bm{w}_k)^T(\bm{w}_{k+1} - \bm{w}_k)\\
&+\frac{L}{2}\|\bm{w}_{k+1} - \bm{w}_k\|^2.
\end{aligned}
\end{equation}
With $\bm{v}_{k+1} = \bm{v}_k + \eta_k \nabla_{\bm{v}} \mathcal{L}(\bm{u}_k,\bm{v}_k,\bm{w}_k; \xi_k)$ , we have
\begin{equation}
\begin{aligned}
       &\mathcal{L}(\bm{u}_{k+1},\bm{v}_{k+1},\bm{w}_{k+1})\\
  \leq &\mathcal{L}(\bm{u}_k,\bm{v}_k,\bm{w}_k) + \eta_k \nabla_{\bm{v}} \mathcal{L}(\bm{u}_k,\bm{v}_k,\bm{w}_k)^\top \nabla_{\bm{v}} \mathcal{L}(\bm{u}_k,\bm{v}_k,\bm{w}_k; \xi_k)  \\
       &+ \frac{L}{2} \eta_k^2\| \nabla_{\bm{v}} \mathcal{L}(\bm{u}_k,\bm{v}_k,\bm{w}_k; \xi_k)\|^2 +\mathcal{R}_k(\bm{u},\bm{w}).
\end{aligned}
\end{equation}
Taking the expectation of the stochastic gradient, we have
\begin{equation}
\label{inequality with expection of gradient v}
\begin{aligned}
     &\mathcal{L}(\bm{u}_{k+1},\bm{v}_{k+1},\bm{w}_{k+1})-\mathcal{L}(\bm{u}_k,\bm{v}_k,\bm{w}_k) \\
     \leq &\frac{L}{2} \eta_k^2 \mathbb{E}[\| \nabla_{\bm{v}} \mathcal{L}(\bm{u}_k,\bm{v}_k,\bm{w}_k; \xi_k) \|^2]\\
     &+ \eta_k \nabla_{\bm{v}} \mathcal{L}(\bm{u}_k,\bm{v}_k,\bm{w}_k)^\top \mathbb{E}[\nabla_{\bm{v}} \mathcal{L}(\bm{u}_k,\bm{v}_k,\bm{w}_k; \xi_k)] \\
     &+ \mathbb{E}[\mathcal{R}_k(\bm{u}, \bm{w})]. 
\end{aligned}
\end{equation}
According to the assumption of the unbiased estimation of the stochastic gradient and defining the variance of the stochastic gradient of $\bm{v}$ as $\sigma_{\bm{v}}$, we have
\begin{equation}
\begin{aligned}
    \mathbb{E}[\nabla_{\bm{u}} \mathcal{L}(\bm{u}_k, \bm{v}_k, \bm{w}_k; \xi_k)] &= \nabla_{\bm{u}} \mathcal{L}(\bm{u}_k, \bm{v}_k, \bm{w}_k)\\
    \mathbb{E}[\nabla_{\bm{v}} \mathcal{L}(\bm{u}_k, \bm{v}_k, \bm{w}_k; \xi_k)] &= \nabla_{\bm{v}} \mathcal{L}(\bm{u}_k, \bm{v}_k, \bm{w}_k),
\end{aligned}
\end{equation}
\begin{equation}
\begin{aligned}
    &\mathbb{E}[\|\nabla_{\bm{v}} \mathcal{L}(\bm{u}_k, \bm{v}_k, \bm{w}_k; \xi_k)\|^2]\\
    =&\mathbb{E}[\|(\nabla_{\bm{v}} \mathcal{L}(\bm{u}_k, \bm{v}_k, \bm{w}_k; \xi_k)-\nabla_{\bm{v}} \mathcal{L}(\bm{u}_k, \bm{v}_k, \bm{w}_k))\\
    &+\nabla_{\bm{v}} \mathcal{L}(\bm{u}_k, \bm{v}_k, \bm{w}_k)\|^2]\\
    \leq& \sigma_{\bm{v}}^2 + \mathbb{E}[\|\nabla_{\bm{v}} \mathcal{L}(\bm{u}_k, \bm{v}_k, \bm{w}_k)\|^2].
\end{aligned}
\end{equation}
Thus, by summing up (\ref{inequality with expection of gradient v}) for $T$ steps, we have the following:
\begin{equation}
\label{inequality summed over k with v}
\begin{aligned}
     &\mathcal{L}(\bm{u}_{T+1},\bm{v}_{T+1},\bm{w}_{T+1})-\mathcal{L}(\bm{u}_1,\bm{v}_1,\bm{w}_1) \\
     \leq &\sum_{k=1}^T(\frac{L}{2} \eta_k^2+\eta_k)\mathbb{E}[\|\nabla_{\bm{v}} \mathcal{L}(\bm{u}_k, \bm{v}_k, \bm{w}_k)\|^2]\\
     &+\frac{L}{2} \eta_k^2\sigma_{\bm{v}}^2+\sum_{k=1}^T \mathbb{E}[\mathcal{R}_k(\bm{u},\bm{w})].
\end{aligned}
\end{equation}
Taking stepsize $\eta_k=\frac{\eta}{\sqrt{k}}$, we have
\begin{equation}
\begin{aligned}
     &\mathcal{L}(\bm{u}_{T+1},\bm{v}_{T+1},\bm{w}_{T+1})-\mathcal{L}(\bm{u}_1,\bm{v}_1,\bm{w}_1) \\
     \leq &\sum_{k=1}^T \left(\frac{L\eta^2}{2k} + \frac{\eta}{\sqrt{k}}\right) \mathbb{E}[\|\nabla_{\bm{u}} \mathcal{L}(\bm{u}_k, \bm{v}_k, \bm{w}_k)\|^2]\\
     &+ \frac{L\eta^2}{2} \sum_{k=1}^T \frac{\sigma_{\bm{v}}^2}{k} + \sum_{k=1}^T \mathbb{E}[\mathcal{R}_k(\bm{u},\bm{w})],
\end{aligned}
\end{equation}

\begin{equation}
\begin{aligned}
    &\sum_{k=1}^T \mathbb{E}[\mathcal{R}_k(\bm{u},\bm{w})]\\
    =&\frac{LT}{2}\mathbb{E}[\|\bm{w}_{k+1} - \bm{w}_k\|^2]
     \\
     &+\sum_{k=1}^T \left(\frac{L\eta^2}{2k} - \frac{\eta}{\sqrt{k}}\right)\mathbb{E}[\|\nabla_{\bm{u}} \mathcal{L}(\bm{u}_k, \bm{v}_k, \bm{w}_k)\|^2]\\
     &+\sum_{k=1}^T\mathbb{E}[\nabla_{\bm{w}} \mathcal{L}(\bm{v}_k, \bm{w}_k, \bm{w}_k)^T(\bm{w}_{k+1} - \bm{w}_k)].
\end{aligned}
\end{equation}
With $\sum_{k=1}^T \frac{1}{k}\approx \log(T), \sum_{k=1}^T \frac{1}{\sqrt{k}}\approx 2\sqrt{T}$, we have
\begin{equation}
    \begin{aligned}
        \frac{L\eta^2}{2k} - \frac{\eta}{\sqrt{k}} &\approx \frac{L\eta^2\log(T)}{2} - 2\eta\sqrt{T},\\
       \frac{L\eta^2}{2k} + \frac{\eta}{\sqrt{k}} &\approx \frac{L\eta^2\log(T)}{2} + 2\eta\sqrt{T}.
    \end{aligned}
\end{equation}
Thus, we have the following:
\begin{equation}
    \begin{aligned}
        &\frac{1}{T}\sum_{k=1}^T\mathbb{E}[\|\nabla_{\bm{v}} \mathcal{L}(\bm{u}_k, \bm{v}_k, \bm{w}_k)\|^2]\\
        \leq &  \frac{\mathcal{L}(\bm{u}_1,\bm{v}_1,\bm{w}_1)-\mathcal{L}(\bm{u}_{T+1},\bm{v}_{T+1},\bm{w}_{T+1})}{\mathcal{C}'_{\eta,L}(T)}\\
        &+\frac{L\eta^2\sigma_{\bm{v}}^2\log(T)}{\mathcal{C}'_{\eta,L}(T)}+\frac{LT\mathbb{E}[\|\bm{w}_{k+1} - \bm{w}_k\|^2]}{2\mathcal{C}'_{\eta,L}(T)}\\
        &+\frac{(\frac{L\eta^2\log(T)}{2}-2\eta\sqrt{T})\mathbb{E}[\|\nabla_{\bm{u}} \mathcal{L}(\bm{u}_k, \bm{v}_k, \bm{w}_k)\|^2]}{\mathcal{C}'_{\eta,L}(T)}\\
        &+\frac{\sum_{k=1}^T\mathbb{E}[\nabla_{\bm{w}} \mathcal{L}(\bm{v}_k, \bm{w}_k, \bm{w}_k)^T(\bm{w}_{k+1} - \bm{w}_k)]}{\mathcal{C}'_{\eta,L}(T)},
    \end{aligned}
\end{equation}
where $\mathcal{C}'_{\eta,L}(T) = T(2\eta\sqrt{T}+\frac{L\eta^2\log(T)}{2})$.

With the assumption that the gradient is bounded, such that $\|\nabla_{\bm{u}} \mathcal{L}(\bm{u}_k, \bm{v}_k, \bm{w}_k)\|\leq G_{\bm{u}}$ and $\|\nabla_{\bm{w}} \mathcal{L}(\bm{u}_k, \bm{v}_k, \bm{w}_k)\|\leq G_{\bm{w}}$, we have the following:
\begin{equation}
\begin{aligned}
        &\frac{(2\eta\sqrt{T}-\frac{L\eta^2\log(T)}{2})\mathbb{E}[\|\nabla_{\bm{u}} \mathcal{L}(\bm{u}_k, \bm{v}_k, \bm{w}_k)\|^2]}{\mathcal{C}'_{\eta,L}(T)}\\
        \leq &   \frac{(\frac{L\eta^2\log(T)-2\eta\sqrt{T}}{2})G_{\bm{u}}^2}{\mathcal{C}'_{\eta,L}(T)},\\
        &\frac{\sum_{k=1}^T\nabla_{\bm{w}} \mathcal{L}(\bm{v}_k, \bm{w}_k, \bm{w}_k)^T(\bm{w}_{k+1} - \bm{w}_k)}{\mathcal{C}'_{\eta,L}(T)}\\
        \leq & \frac{G_{\bm{w}}\sum_{k=1}^T \mathbb{E}[\|\bm{w}_{k+1} - \bm{w}_k\|]}{\mathcal{C}'_{\eta,L}(T)}.
\end{aligned}
\end{equation}
Since $\bm{w}_k$ is convergent, then $ \mathbb{E}[\|\bm{w}_{k+1} - \bm{w}_k\|]$ is bounded, assuming $\mathbb{E}[\|\bm{w}_{k+1} - \bm{w}_k\|\leq M_{\bm{w}}$, we have the following.
\begin{equation}
    \begin{aligned}
        &\frac{1}{T}\sum_{k=1}^T\mathbb{E}[\|\nabla_{\bm{v}} \mathcal{L}(\bm{u}_k, \bm{v}_k, \bm{w}_k)\|^2]\\
        \leq &  \frac{\mathcal{L}(\bm{u}_1,\bm{v}_1,\bm{w}_1)-\mathcal{L}(\bm{u}_{T+1},\bm{v}_{T+1},\bm{w}_{T+1})}{\mathcal{C}'_{\eta,L}(T)}+\frac{L\eta^2\sigma_{\bm{v}}^2\log(T)}{\mathcal{C}'_{\eta,L}(T)}\\
        &+\frac{LTM_{\bm{w}}^2}{2\mathcal{C}'_{\eta,L}(T)}+\frac{(\frac{L\eta^2\log(T)}{2}-2\eta\sqrt{T})G_{\bm{u}}}{\mathcal{C}'_{\eta,L}(T)}+\frac{G_{\bm{w}}TM_{\bm{w}}}{\mathcal{C}'_{\eta,L}(T)}
    \end{aligned}
\end{equation}
Since $\mathcal{C}'_{\eta,L}(T) = T(2\eta\sqrt{T}+\frac{L\eta^2\log(T)}{2})=\mathcal{O}(T^{1.5})$, we have
\begin{equation}
\begin{aligned}
    &\max\{\frac{1}{\mathcal{C}'_{\eta,L}(T)},\frac{L\eta^2\sigma_{\bm{v}}^2\log(T)}{\mathcal{C}'_{\eta,L}(T)},\frac{LTM_{\bm{w}}^2}{2\mathcal{C}'_{\eta,L}(T)},\\
    &\frac{(\frac{L\eta^2\log(T)}{2}-2\eta\sqrt{T})G_{\bm{u}}}{\mathcal{C}'_{\eta,L}(T)},\frac{G_{\bm{w}}TM_{\bm{w}}}{\mathcal{C}'_{\eta,L}(T)}\}\\
    = &\mathcal{O}(\frac{1}{\sqrt{T}}).
\end{aligned}
\end{equation}
Thus, $\bm{v}$ has a convergence rate of $\mathcal{O}(\frac{1}{\sqrt{T}})$.
\twocolumn
\bibliographystyle{IEEEtran}  
\bibliography{reference}     

\begin{thebibliography}{10}
\providecommand{\url}[1]{#1}
\csname url@samestyle\endcsname
\providecommand{\newblock}{\relax}
\providecommand{\bibinfo}[2]{#2}
\providecommand{\BIBentrySTDinterwordspacing}{\spaceskip=0pt\relax}
\providecommand{\BIBentryALTinterwordstretchfactor}{4}
\providecommand{\BIBentryALTinterwordspacing}{\spaceskip=\fontdimen2\font plus
\BIBentryALTinterwordstretchfactor\fontdimen3\font minus \fontdimen4\font\relax}
\providecommand{\BIBforeignlanguage}[2]{{%
\expandafter\ifx\csname l@#1\endcsname\relax
\typeout{** WARNING: IEEEtran.bst: No hyphenation pattern has been}%
\typeout{** loaded for the language `#1'. Using the pattern for}%
\typeout{** the default language instead.}%
\else
\language=\csname l@#1\endcsname
\fi
#2}}
\providecommand{\BIBdecl}{\relax}
\BIBdecl

\bibitem{li2023joint}
Y.~Li, Z.~Wei, and Z.~Feng, ``Joint subcarrier and power allocation for uplink integrated sensing and communication system,'' \emph{IEEE Sensors Journal}, 2023.

\bibitem{montoliu2020indoor}
R.~Montoliu, E.~Sansano, A.~Gasc{\'o}, O.~Belmonte, and A.~Caballer, ``Indoor positioning for monitoring older adults at home: Wi-fi and ble technologies in real scenarios,'' \emph{Electronics}, vol.~9, no.~5, p. 728, 2020.

\bibitem{bao2022wi}
N.~Bao, J.~Du, C.~Wu, D.~Hong, J.~Chen, R.~Nowak, and Z.~Lv, ``Wi-breath: A wifi-based contactless and real-time respiration monitoring scheme for remote healthcare,'' \emph{IEEE journal of biomedical and health informatics}, 2022.

\bibitem{jin2018whole}
Y.~Jin, Z.~Tian, M.~Zhou, Z.~Li, and Z.~Zhang, ``A whole-home level intrusion detection system using wifi-enabled iot,'' in \emph{2018 14th International Wireless Communications \& Mobile Computing Conference (IWCMC)}.\hskip 1em plus 0.5em minus 0.4em\relax IEEE, 2018, pp. 494--499.

\bibitem{fu2018writing}
Z.~Fu, J.~Xu, Z.~Zhu, A.~X. Liu, and X.~Sun, ``Writing in the air with wifi signals for virtual reality devices,'' \emph{IEEE Transactions on Mobile Computing}, vol.~18, no.~2, pp. 473--484, 2018.

\bibitem{liu2015rss}
C.~Liu, D.~Fang, Z.~Yang, H.~Jiang, X.~Chen, W.~Wang, T.~Xing, and L.~Cai, ``Rss distribution-based passive localization and its application in sensor networks,'' \emph{IEEE Transactions on Wireless Communications}, vol.~15, no.~4, pp. 2883--2895, 2015.

\bibitem{gu2019wifi}
Y.~Gu, X.~Zhang, Z.~Liu, and F.~Ren, ``Wifi-based real-time breathing and heart rate monitoring during sleep,'' in \emph{2019 IEEE Global Communications Conference (GLOBECOM)}.\hskip 1em plus 0.5em minus 0.4em\relax IEEE, 2019, pp. 1--6.

\bibitem{wang2020csi}
X.~Wang, C.~Yang, and S.~Mao, ``On csi-based vital sign monitoring using commodity wifi,'' \emph{ACM Transactions on Computing for Healthcare}, vol.~1, no.~3, pp. 1--27, 2020.

\bibitem{xie2024robust}
X.~Xie, D.~Zhang, Y.~Li, Y.~Hu, Q.~Sun, and Y.~Chen, ``Robust wifi respiration sensing in the presence of interfering individual,'' \emph{IEEE Transactions on Mobile Computing}, 2024.

\bibitem{kontou2023contactless}
P.~Kontou, S.~B. Smida, and D.~E. Anagnostou, ``Contactless respiration monitoring using wi-fi and artificial neural network detection method,'' \emph{IEEE Journal of Biomedical and Health Informatics}, 2023.

\bibitem{ngamakeur2023passive}
K.~Ngamakeur, S.~Yongchareon, J.~Yu, and S.~Islam, ``Passive infrared sensor dataset and deep learning models for device-free indoor localization and tracking,'' \emph{Pervasive and Mobile Computing}, vol.~88, p. 101721, 2023.

\bibitem{yan2021device}
J.~Yan, L.~Wan, W.~Wei, X.~Wu, W.-P. Zhu, and D.~P.-K. Lun, ``Device-free activity detection and wireless localization based on cnn using channel state information measurement,'' \emph{IEEE Sensors Journal}, vol.~21, no.~21, pp. 24\,482--24\,494, 2021.

\bibitem{wang2018low}
J.~Wang, J.~Xiong, H.~Jiang, K.~Jamieson, X.~Chen, D.~Fang, and C.~Wang, ``Low human-effort, device-free localization with fine-grained subcarrier information,'' \emph{IEEE Transactions on Mobile Computing}, vol.~17, no.~11, pp. 2550--2563, 2018.

\bibitem{wang2018quasi}
D.~Wang, M.~Fattouche, F.~M. Ghannouchi, and X.~Zhan, ``Quasi-optimal subcarrier selection dedicated for localization with multicarrier-based signals,'' \emph{IEEE Systems Journal}, vol.~13, no.~2, pp. 1157--1168, 2018.

\bibitem{zhang2022hybrid}
R.~Zhang, Z.~Wang, G.~Li, Y.~Wang, J.~Shuai, J.~Zheng, J.~Huang, and S.~Bao, ``Hybrid subcarrier selection method for vital sign monitoring with long-term and short-term data considerations,'' \emph{IEEE Sensors Journal}, vol.~22, no.~23, pp. 23\,209--23\,220, 2022.

\bibitem{liu2021investigating}
R.~Liu, J.~Gao, J.~Zhang, D.~Meng, and Z.~Lin, ``Investigating bi-level optimization for learning and vision from a unified perspective: A survey and beyond,'' \emph{IEEE Transactions on Pattern Analysis and Machine Intelligence}, vol.~44, no.~12, pp. 10\,045--10\,067, 2021.

\bibitem{zhang2022revisiting}
Y.~Zhang, G.~Zhang, P.~Khanduri, M.~Hong, S.~Chang, and S.~Liu, ``Revisiting and advancing fast adversarial training through the lens of bi-level optimization,'' in \emph{International Conference on Machine Learning}.\hskip 1em plus 0.5em minus 0.4em\relax PMLR, 2022, pp. 26\,693--26\,712.

\bibitem{chen2022gradient}
C.~Chen, X.~Chen, C.~Ma, Z.~Liu, and X.~Liu, ``Gradient-based bi-level optimization for deep learning: A survey,'' \emph{arXiv preprint arXiv:2207.11719}, 2022.

\bibitem{miao2016genetic}
C.~Miao, G.~Du, Y.~Xia, and D.~Wang, ``Genetic algorithm for mixed integer nonlinear bilevel programming and applications in product family design,'' \emph{Mathematical Problems in Engineering}, vol. 2016, no.~1, p. 1379315, 2016.

\bibitem{tahernejad2020branch}
S.~Tahernejad, T.~K. Ralphs, and S.~T. DeNegre, ``A branch-and-cut algorithm for mixed integer bilevel linear optimization problems and its implementation,'' \emph{Mathematical Programming Computation}, vol.~12, no.~4, pp. 529--568, 2020.

\bibitem{tang2016class}
Y.~Tang, J.-P.~P. Richard, and J.~C. Smith, ``A class of algorithms for mixed-integer bilevel min--max optimization,'' \emph{Journal of Global Optimization}, vol.~66, pp. 225--262, 2016.

\bibitem{avraamidou2017mixed}
S.~Avraamidou, N.~Diangelakis, and E.~Pistikopoulos, ``Mixed integer bilevel optimization through multi-parametric programming,'' \emph{Foundations of computer aided process operations/chemical process control}, pp. In--Press, 2017.

\bibitem{dumouchelle2024neur2bilo}
J.~Dumouchelle, E.~Julien, J.~Kurtz, and E.~Khalil, ``Neur2bilo: Neural bilevel optimization,'' \emph{Advances in Neural Information Processing Systems}, vol.~37, pp. 86\,688--86\,719, 2024.

\bibitem{jiao2022asynchronous}
Y.~Jiao, K.~Yang, T.~Wu, D.~Song, and C.~Jian, ``Asynchronous distributed bilevel optimization,'' in \emph{The Eleventh International Conference on Learning Representations}, 2022.

\bibitem{jiao2022distributed}
Y.~Jiao, K.~Yang, and D.~Song, ``Distributed distributionally robust optimization with non-convex objectives,'' \emph{Advances in neural information processing systems}, vol.~35, pp. 7987--7999, 2022.

\bibitem{huang2024triadic}
Y.~Huang, K.~Yang, Z.~Zhu, and L.~Chen, ``Triadic-ocd: Asynchronous online change detection with provable robustness, optimality, and convergence,'' in \emph{International Conference on Machine Learning}.\hskip 1em plus 0.5em minus 0.4em\relax PMLR, 2024, pp. 20\,382--20\,412.

\bibitem{chen2024robust}
X.~Chen, Y.~Xiong, and K.~Yang, ``Robust beamforming for downlink multi-cell systems: A bilevel optimization perspective,'' in \emph{Proceedings of the AAAI Conference on Artificial Intelligence}, vol.~38, no.~8, 2024, pp. 7969--7977.

\bibitem{jian2024tri}
C.~Jian, K.~Yang, and Y.~Jiao, ``Tri-level navigator: Llm-empowered tri-level learning for time series ood generalization,'' \emph{Advances in Neural Information Processing Systems}, vol.~37, pp. 110\,613--110\,642, 2024.

\bibitem{yang2013rssi}
Z.~Yang, Z.~Zhou, and Y.~Liu, ``From rssi to csi: Indoor localization via channel response,'' \emph{ACM Computing Surveys (CSUR)}, vol.~46, no.~2, pp. 1--32, 2013.

\bibitem{wu2022wifi}
D.~Wu, Y.~Zeng, F.~Zhang, and D.~Zhang, ``Wifi csi-based device-free sensing: from fresnel zone model to csi-ratio model,'' \emph{CCF Transactions on Pervasive Computing and Interaction}, pp. 1--15, 2022.

\bibitem{yan2019wiact}
H.~Yan, Y.~Zhang, Y.~Wang, and K.~Xu, ``Wiact: A passive wifi-based human activity recognition system,'' \emph{IEEE Sensors Journal}, vol.~20, no.~1, pp. 296--305, 2019.

\bibitem{sharma2022attention}
P.~Sharma, Z.~Zhang, T.~B. Conroy, X.~Hui, and E.~C. Kan, ``Attention detection by heartbeat and respiratory features from radio-frequency sensor,'' \emph{Sensors}, vol.~22, no.~20, p. 8047, 2022.

\bibitem{fan2024contactless}
D.~Fan, X.~Yang, N.~Zhao, L.~Guan, M.~M. Arslan, M.~Ullah, M.~A. Imran, and Q.~H. Abbasi, ``A contactless breathing pattern recognition system using deep learning and wifi signal,'' \emph{IEEE Internet of Things Journal}, 2024.

\bibitem{graefenstein2009wireless}
J.~Graefenstein, A.~Albert, P.~Biber, and A.~Schilling, ``Wireless node localization based on rssi using a rotating antenna on a mobile robot,'' in \emph{2009 6Th Workshop on positioning, navigation and communication}.\hskip 1em plus 0.5em minus 0.4em\relax IEEE, 2009, pp. 253--259.

\bibitem{matricardi2023performance}
E.~Matricardi, L.~Pucci, E.~Paolini, W.~Xu, and A.~Giorgetti, ``Performance analysis of a multistatic joint sensing and communication system,'' in \emph{2023 IEEE 34th Annual International Symposium on Personal, Indoor and Mobile Radio Communications (PIMRC)}.\hskip 1em plus 0.5em minus 0.4em\relax IEEE, 2023, pp. 1--6.

\bibitem{zhang2022wi}
R.~Zhang, C.~Jiang, S.~Wu, Q.~Zhou, X.~Jing, and J.~Mu, ``Wi-fi sensing for joint gesture recognition and human identification from few samples in human-computer interaction,'' \emph{IEEE Journal on Selected Areas in Communications}, vol.~40, no.~7, pp. 2193--2205, 2022.

\bibitem{patwari2013breathfinding}
N.~Patwari, L.~Brewer, Q.~Tate, O.~Kaltiokallio, and M.~Bocca, ``Breathfinding: A wireless network that monitors and locates breathing in a home,'' \emph{IEEE Journal of Selected Topics in Signal Processing}, vol.~8, no.~1, pp. 30--42, 2013.

\bibitem{hillyard2018experience}
P.~Hillyard, A.~Luong, A.~S. Abrar, N.~Patwari, K.~Sundar, R.~Farney, J.~Burch, C.~Porucznik, and S.~H. Pollard, ``Experience: Cross-technology radio respiratory monitoring performance study,'' in \emph{Proceedings of the 24th Annual International Conference on Mobile Computing and Networking}, 2018, pp. 487--496.

\bibitem{gu2018your}
Y.~Gu, X.~Zhang, C.~Li, F.~Ren, J.~Li, and Z.~Liu, ``Your wifi knows how you behave: Leveraging wifi channel data for behavior analysis,'' in \emph{2018 IEEE Global Communications Conference (GLOBECOM)}.\hskip 1em plus 0.5em minus 0.4em\relax IEEE, 2018, pp. 1--6.

\bibitem{wang2020resilient}
X.~Wang, C.~Yang, and S.~Mao, ``Resilient respiration rate monitoring with realtime bimodal csi data,'' \emph{IEEE Sensors Journal}, vol.~20, no.~17, pp. 10\,187--10\,198, 2020.

\bibitem{fahad2025ensemble}
N.~Fahad, M.~Touhiduzzaman, and E.~Bulut, ``Ensemble learning based wifi sensing using spatially distributed tx-rx links,'' in \emph{IEEE International Conference on Computing, Networking and Communications (ICNC), Honolulu, Hawaii, USA}, 2025.

\bibitem{chu2021wifi}
F.-Y. Chu, C.-J. Chiu, A.-H. Hsiao, K.-T. Feng, and P.-H. Tseng, ``Wifi csi-based device-free multi-room presence detection using conditional recurrent network,'' in \emph{2021 IEEE 93rd Vehicular Technology Conference (VTC2021-Spring)}.\hskip 1em plus 0.5em minus 0.4em\relax IEEE, 2021, pp. 1--5.

\bibitem{biswas2019literature}
A.~Biswas and C.~Hoyle, ``A literature review: solving constrained non-linear bi-level optimization problems with classical methods,'' in \emph{International Design Engineering Technical Conferences and Computers and Information in Engineering Conference}, vol. 59193.\hskip 1em plus 0.5em minus 0.4em\relax American Society of Mechanical Engineers, 2019, p. V02BT03A025.

\bibitem{denegre2011interdiction}
S.~DeNegre, \emph{Interdiction and discrete bilevel linear programming}.\hskip 1em plus 0.5em minus 0.4em\relax Lehigh University, 2011.

\bibitem{soares2023deterministic}
I.~Soares, M.~J. Alves, and C.~H. Antunes, ``A deterministic bounding algorithm vs. a hybrid meta-heuristic to deal with a bilevel mixed-integer nonlinear optimization model for electricity dynamic pricing,'' \emph{Computers \& Operations Research}, vol. 155, p. 106195, 2023.

\bibitem{beykal2020domino}
B.~Beykal, S.~Avraamidou, I.~P. Pistikopoulos, M.~Onel, and E.~N. Pistikopoulos, ``Domino: Data-driven optimization of bi-level mixed-integer nonlinear problems,'' \emph{Journal of Global Optimization}, vol.~78, pp. 1--36, 2020.

\bibitem{yiu2015gaussian}
S.~Yiu and K.~Yang, ``Gaussian process assisted fingerprinting localization,'' \emph{IEEE Internet of Things Journal}, vol.~3, no.~5, pp. 683--690, 2015.

\bibitem{chen2016robust}
L.~Chen, K.~Yang, and X.~Wang, ``Robust cooperative wi-fi fingerprint-based indoor localization,'' \emph{IEEE Internet of Things Journal}, vol.~3, no.~6, pp. 1406--1417, 2016.

\bibitem{yang2008distributed}
K.~Yang, Y.~Wu, J.~Huang, X.~Wang, and S.~Verd{\'u}, ``Distributed robust optimization for communication networks,'' in \emph{IEEE INFOCOM 2008-The 27th Conference on Computer Communications}.\hskip 1em plus 0.5em minus 0.4em\relax IEEE, 2008, pp. 1157--1165.

\bibitem{yang2014distributed}
K.~Yang, J.~Huang, Y.~Wu, and M.~Chiang, ``Distributed robust optimization (dro), part i: framework and example,'' \emph{Optimization and Engineering}, vol.~15, no.~1, pp. 35--67, 2014.

\bibitem{ghadimi2020single}
S.~Ghadimi, A.~Ruszczynski, and M.~Wang, ``A single timescale stochastic approximation method for nested stochastic optimization,'' \emph{SIAM Journal on Optimization}, vol.~30, no.~1, pp. 960--979, 2020.

\bibitem{huang2024wimans}
S.~Huang, K.~Li, D.~You, Y.~Chen, A.~Lin, S.~Liu, X.~Li, and J.~A. McCann, ``Wimans: A benchmark dataset for wifi-based multi-user activity sensing,'' in \emph{European Conference on Computer Vision}.\hskip 1em plus 0.5em minus 0.4em\relax Springer, 2024, pp. 72--91.

\bibitem{rao2023mffaloc}
X.~Rao, Z.~Luo, Y.~Luo, Y.~Yi, G.~Lei, and Y.~Cao, ``Mffaloc: Csi-based multi-features fusion adaptive device-free passive indoor fingerprinting localization,'' \emph{IEEE Internet of Things Journal}, 2023.

\bibitem{he2023robust}
Z.~He, X.~Zhang, Y.~Wang, Y.~Lin, G.~Gui, and H.~Gacanin, ``A robust csi-based wi-fi passive sensing method using attention mechanism deep learning,'' \emph{IEEE Internet of Things Journal}, 2023.

\bibitem{yang2022efficientfi}
J.~Yang, X.~Chen, H.~Zou, D.~Wang, Q.~Xu, and L.~Xie, ``Efficientfi: Toward large-scale lightweight wifi sensing via csi compression,'' \emph{IEEE Internet of Things Journal}, vol.~9, no.~15, pp. 13\,086--13\,095, 2022.

\end{thebibliography}
\end{document}